\documentclass{article}

\usepackage[english]{babel}

\usepackage[letterpaper,top=2cm,bottom=2cm,left=3cm,right=3cm,marginparwidth=1.75cm]{geometry}

\usepackage[table,xcdraw]{xcolor}
\usepackage[normalem]{ulem}
\useunder{\uline}{\ul}{}
\usepackage{longtable}
\usepackage{lscape}
\usepackage{setspace}
\newcommand*{\at }{@}
\usepackage{caption}

\usepackage{amsmath}
\usepackage{graphicx}
\usepackage[colorlinks=true, allcolors=blue]{hyperref}
\usepackage{mathptmx}
\usepackage{booktabs}
\usepackage{textcomp}
\usepackage[T1]{fontenc}
\usepackage[sort,comma,authoryear,round]{natbib}

\title{The TEAMx-PC22 Alpine field campaign -- Objectives, instrumentation, and observed phenomena}
\author{Lena Pfister$^1$, Alexander Gohm$^1$, Meinolf Kossmann$^2$, Andreas Wieser$^3$, Nevio Babić$^{3,7}$, \\ Jan Handwerker$^3$, Norman Wildmann$^4$, Hannes Vogelmann$^5$, Kathrin Baumann-Stanzer$^6$, \\ Almut Alexa$^{1,4}$, Karl Lapo$^1$, Ivan Paunović$^2$, Ronny Leinweber$^2$, Katrin Seldmeier$^2$, \\ Manuela Lehner$^1$, Alexander Hieden$^6$, Johannes Speidel$^5$, Maria Federer$^5$, Mathias W. Rotach$^1$
}

\begin{document}
\maketitle
This manuscript was submitted on 07. Nov 2023 and is under review at Meteorologische Zeitschrift - Contributions to Atmospheric Sciences, Borntraeger Science Publishers, Germany (ISSN: 0941-2948; ISSN online version: 1610-1227)
\begin{itemize} \itemsep-2mm
\item[$^1$]University of Innsbruck, Department of Atmospheric and Cryospheric Sciences, Innsbruck, Austria 
\item[$^2$]Deutscher Wetterdienst, Germany 
\item[$^3$]Karlsruhe Institute of Technology (KIT), Institute of Meteorology and Climate Research - Department Troposphere Research (IMK-TRO), Karlsruhe, Germany 
\item[$^4$]Deutsches Zentrum für Luft- und Raumfahrt e.V., Institut für Physik der Atmosphäre, Oberpfaffenhofen, Germany
\item[$^5$]Karlsruhe Institute of Technology (KIT), Institute of Meteorology and Climate Research - Atmospheric Environmental Research (IMK-IFU), Garmisch-Partenkirchen, Germany
\item[$^6$]GeoSphere Austria - Bundesanstalt für Geologie, Geophysik, Klimatologie und Meteorologie, Vienna, Austria 
\item[$^7$]Croatia Control Ltd., Zagreb, Croatia
\end{itemize}
Correspondence: \\
Lena Pfister, University of Innsbruck, Department of Atmospheric and Cryospheric Sciences, Innsbruck, Austria\\
lena.pfister@uibk.ac.at

\begin{abstract}
The multi-scale transport and exchange processes in the atmosphere over mountains---programme and experiment (TEAMx) wants to advance the understanding of transport and exchange processes over mountainous terrain as well as to collect unique multi-scale datasets that can be used, e.g., for process studies, model development and model evaluation. The TEAMx Observational Campaign (TOC) is planned to take place between 2024 and 2025. In summer 2022 a TEAMx pre-campaign (TEAMx-PC22) was conducted in the Inn Valley and one of its tributaries, the Weer Valley, to test the suitability and required logistics of measurement sites, to evaluate their value for the main campaign, and to test new observation techniques in complex terrain. Scientifically, this campaign focused on resolving the mountain boundary layer and valley wind systems on multiple scales. Through the combined effort of six institutions the pre-campaign can be deemed successful. A detailed description of the setup at each sub-target area is given. Due to the spatial distribution of instruments and their spatio-temporal resolution, atmospheric processes and phenomena like valley winds have been investigated at different locations and on different scales. Furthermore, scale interactions were detected and are discussed in detail in two example cases. Additionally, observational gaps were determined which should be closed for the TOC. Data of the pre-campaign are publicly available online and can be used for process studies, demonstrating the utility of new observation methods, model verification, and for data assimilation.
\end{abstract}
\section{Introduction}
\subsection{Motivation and objectives for TEAMx}
A key characteristic of the atmospheric boundary layer is its role in the exchange of heat, momentum, moisture, and other components between the earth’s surface and the atmosphere. In contrast to flat terrain, where this exchange is mainly the result of vertical turbulent mixing, a number of different processes contribute to the total transport and exchange in the mountain boundary layer (MoBL), spanning a range of different scales \citep{Lehner2018}. Examples include the transport by thermally driven slope and valley winds \citep{Giovannini2020, Serafin2018}, through forced lifting by the topography \citep{Kirshbaum2018}, the turbulent exchange resulting from orographically induced gravity waves \citep{Vosper2018}, and from horizontal wind shear produced by thermally driven circulations \citep{Goger2018}. While the impact of mountains on weather and climate has been studied for many decades, including major coordinated field campaigns such as the Alpine Experiment (ALPEX; \citealt{wmo-86Aag}), the Pyrénées Experiment (PYREX; \citealt{bougeault1997pyrex}), the Convective and Orographically-induced Precipitation Study (COPS; \citealt{Wulfmeyer2011}), and the Mesoscale Alpine Programme (MAP; \citealt{Bougeault2001}) in Europe, recent advances in computing power and measurement technologies have made it possible to observe and model previously unresolved scales, which has motivated the mountain meteorology community to initiate a new research program called TEAMx (multi-scale transport and exchange processes in the atmosphere over mountains---programme and experiment\footnote{\url{www.teamx-programme.org}}).

TEAMx is an international programme that aims at (i) advancing our understanding of transport and exchange processes over mountainous terrain; (ii) improving the representation of these processes in numerical weather and climate prediction models; (iii) collecting a unique dataset that can be used for process studies, model evaluation, and the development of model parameterizations; and (iv) communicating the new results to weather and climate service providers to reduce the error and uncertainty of end-user products \citep{Serafin2020, Rotach2022}. These goals will be addressed through a one-year long observational campaign, the so-called TEAMx Observational Campaign (TOC), taking place in the Alps between 2024 and 2025. For the TOC, a number of coordinated modeling activities, such as process-oriented model intercomparison studies, large-eddy simulations, and seamless model runs from global to sub-km scales, and extended near realtime measurement data assimilation will be conducted, as well. The TOC will bring together multiple research groups from around the world to sample the MoBL at different scales. Observations will span across the Alpine ridge, with target areas in the pre-Alpine regions north (Germany) and south (Italy) of the Alps, in inner-Alpine north-south oriented (Adige Valley, Italy) and west-east oriented (Inn Valley, Austria) valleys, and in the crest region (e.g., Sarntal Alps, Italy).

In summer 2022, a small number of research groups conducted a pre-campaign (TEAMx-PC22) in the Inn Valley target area (IVTA) to identify suitable measurement sites and to test measurement strategies for the TOC.
\subsection{Motivation and objectives for the TEAMx pre-campaign 2022}
Several field campaigns have already been carried out jointly by various institutions in the Inn Valley and its tributary valleys. However, these experiments focused on specific flows or were lacking observations on larger scales, making the investigation of scale interaction difficult. For example, the field campaign ``Mesoscale Experiment in the Region of Kufstein and Rosenheim'' (MERKUR) in 1982 was missing continuous observations of the MoBL \citep{Freytag1983}, while the ``Mesoscale Alpine Programme'' (MAP) in 1999 studied only gap flows in the Wipp Valley, which is one of the largest tributaries of the Inn Valley \citep{Mayr2003}. The field campaign ``Air Pollution, Traffic Noise and Related Health Effects in the Alpine Space'' (ALPNAP) in 2006 focused only on winter conditions \citep[][Chapter 7.1]{Heimann2007}, while the ``Penetration and Interruption of Alpine Foehn'' (PIANO) project in 2017 was limited to the investigation of cold pool processes during foehn \citep{Haid2020}. Last but not least, the ``Cross-Valley Flow in the Inn Valley Investigated by Dual-Doppler Lidar Measurements'' (CROSSINN) project in 2019 had cross-valley observations in high spatio-temporal resolution, but was lacking along-valley observations \citep{Adler2021}. The concept of TEAMx-PC22 addresses these limitations by establishing multiple field sites to observe the MoBL at a high spatio-temporal resolution. This strategy helped to resolve characteristics of local flows, such as valley winds, and associated processes in the MoBL on multiple scales and advance our knowledge. 

The Inn Valley target area (IVTA) is suitable for investigating the MoBL as well as processes  on different scales from turbulence to valley winds to foehn. An example is the nocturnal valley exit-jet blowing from the Inn Valley into the Bavarian foreland. These exit-jets play an important role for the transport of air pollutants \citep{Banta1995} but also contribute to the alleviation of urban heat islands in cities located at valley exits during summertime heat waves \citep{Weischet1983}. Previous research primarily focused on the height and strength of these jets and on how far jets can reach into adjacent plains \citep{Pamperin1985, Zangl2004}. Recent developments in Doppler wind lidar technology \citep{Banta2013, Smalikho2017} and large eddy simulations \citep[LES;][]{Chow2019} now offer new possibilities to better understand and simulate nocturnal Inn Valley exit-jets. For investigating not only valley exit-jets, but processes on multiple scales in the IVTA a range of instruments were utilized for TEAMx-PC22, including different types of  remote sensing instrumentation,  (multi-level) point observations, a fleet of uncrewed aerial system (UAS), distributed temperature sensing (DTS), and radiosondes each being capable of resolving different scales. To combine the strength of different instruments and to create a network, the IVTA was divided into several sub-target areas each with a different research focus. 

An additional focus of TEAMx-PC22 was on resolving scale interactions. The interaction between the nocturnal flow in the main valley and the outflow from tributary valleys is one of these processes. The lack of spatio-temporal resolution to adequately observe micro- and mesoscale flow features across the confluence interface of a parent valley and its tributary \citep{Zardi2013} still leaves gaps in our understanding and leads to an incomplete representation in numerical weather prediction (NWP) models.

In summary, during TEAMx-PC22 six institutes combined their efforts to gain experience and test the suitability of four sub-target areas of the Inn Valley. The newest observation techniques as well as required logistics were tested and evaluated to investigate scale interactions within the Alps for the upcoming TEAMx Observational Campaign in 2024 and 2025. This paper will give an overview of the data collected during TEAMx-PC22, highlight some of the observed processes and scale interactions, and discuss the success of the campaign to prepare for the TOC.
\section{Physical setting and instrumentation of sub-target areas}
\subsection{Physical setting}
\begin{figure}
    \centering
    \includegraphics[width = \textwidth]{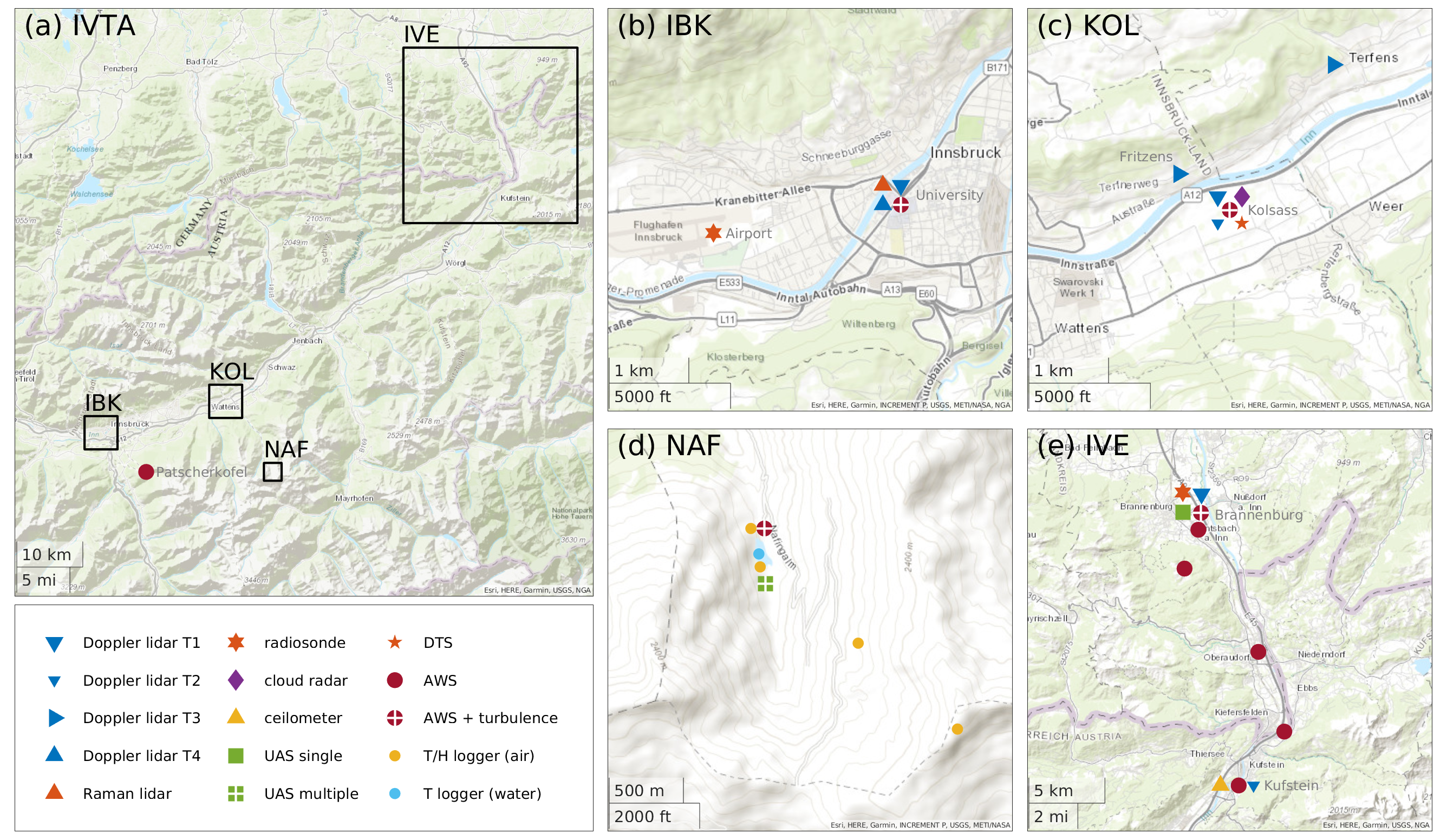}
    \caption{(a) Overview of the Inn Valley target area (IVTA) with several sub-target areas indicated by black squares: (b) Innsbruck (IBK), (c) Kolsass (KOL), (d) Nafingalm (NAF), and (e) Inn Valley Exit (IVE). Instruments in each sub-target area are represented by markers (see legend). Doppler lidars are divided into four different types (T1-T4): long-range (T1) and short-range (T2) lidars performing conical scans, horizontally scanning lidars (T3), and a vertically staring lidar (T4). Automatic weather stations (AWS) with and without ability to measure turbulence and sites with single and multiple UAS are indicated by different markers, respectively. See text for further details on instruments.}
\label{fig1:target}
\end{figure}

\begin{figure}
    \centering
    \includegraphics[width = \textwidth]{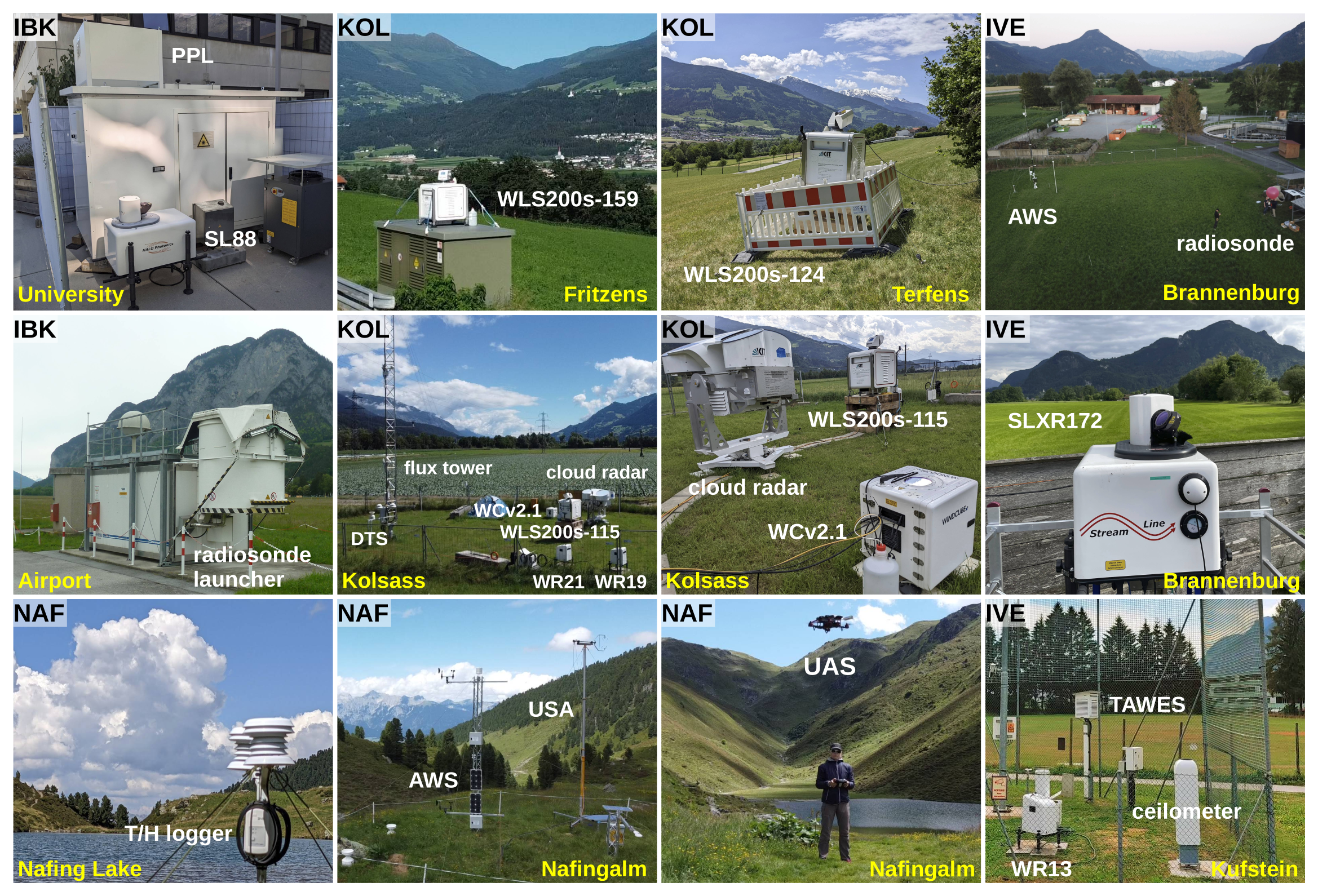}
    \caption{Photo collage of instruments deployed in the Inn Valley target area. The sub-target areas Innsbruck (IBK), Kolsass (KOL), Nafingalm (NAF), and Inn Valley Exit (IVE) are indicated in the upper left corner of each photo (see also Fig.~\ref{fig1:target}). Names of instruments and sub-target area are given in white and yellow font, respectively. See Table~\ref{tab:instruments} for further details on instruments. 
    }
\label{fig2:photos}
\end{figure}

TEAMx-PC22 had instruments running from June until end of September 2022 within the IVTA (Fig.~\ref{fig1:target}), however, not all were running simultaneously. A detailed summary of the equipment within each sub-target area and dates with running instruments is given in Table~\ref{tab:instruments}. Further, a photo collage giving impressions of each sub-target area and its equipment is shown in Figure~\ref{fig2:photos}. 

The main Inn Valley target area (IVTA) is located in the western part of Austria and is oriented approximately southwest--northeast, with a depth of about 2000~m and a width at the valley floor of about 3~km in the sub-target area Innsbruck~(IBK) including the outlet of the Wipp Valley towards the south (Fig.~\ref{fig1:target}). The valley floor in IBK is located at about 580 m above mean sea level (AMSL). The sub-target area Kolsass (KOL) is located roughly 20~km east-northeast of IBK at 550 m AMSL, includes the outlet of the tributary Weer Valley and the valley width at the bottom is about 2~km. The sub-target area Nafingalm (NAF) is located roughly 13~km southeast of KOL at the end of the tributary Weer Valley at a mountain pasture called Nafingalm with a small natural lake called Nafingsee (Nafing Lake) situated at the valley floor at 1920~m AMSL. The head of the valley has a size of approximately 2$\times$2~km$^2$, a difference in altitude of 350~to 500~m, a north--south orientation, an average slope angle of 30\textdegree~to the east, west, and north, and a nearly homogeneous low-level vegetation characterized by a mixture of grass and shrubs. Following the Inn River downvalley towards the Inn Valley Exit (IVE) sub-target area, the valley gets narrower with a width of 1 to 2.3~km at the bottom and a depth of about 600~m. After Kufstein, the orientation changes to south-southeast--north-northwest (cf.~Fig.~\ref{fig1:target}). Each sub-target area is equipped with an automatic weather station (AWS) and with a Doppler wind lidar, with the exception of NAF, creating an along-valley transect of observations. Besides NAF, all AWS stations are operated continuously by the Department of Atmospheric and Cryospheric Sciences of the University of Innsbruck (ACINN), the German Meteorological Service (DWD), and GeoSphere Austria at least until the end of the TOC. A detailed description of the instruments deployed in each sub-target area is given in Table~\ref{tab:instruments} and in the following sections.

\subsection{Sub-target area Innsbruck (IBK) \label{sec:IBK}}
The focus within the sub-target area IBK  (Fig.~\ref{fig1:target}b) was on resolving the vertical structure of the MoBL. The Institute of Meteorology and Climate Research - Atmospheric Environmental Research (IMK-IFU) of the Karlsruhe Institute of Technology (KIT) tested the performance of the new Purple Pulse Raman lidar (PPL) system \citep{Lange2019} and evaluated whether its combination with a Doppler wind lidar is suitable for the calculation of turbulent vertical heat and moisture fluxes over complex terrain in a similar way as it has been done in the past over flat terrain \citep[e.g.][]{Behrendt2020,Lareau2020}. Furthermore, it was of interest, if this setup is also suitable for examining mixing, transport and interaction between the boundary layer and the free troposphere, in particular in the presence of mesoscale to synoptic-scale wind systems (e.g. foehn). It was operated between 09~August and 04~October 2022 at the University of Innsbruck in the forecourt of the Campus Innrain, next to the university building at Innrain~52f where ACINN is situated. PPL measured vertical profiles of water vapor, temperature and aerosol backscatter at an interval of 10~s and with a range-gate length of 3.75~m, typically averaged over 26 range gates to 97.5~m. Its operating wavelength is 355~nm with an average output power of 14~W at 200~Hz.

ACINN supported the efforts of KIT/IMK-IFU with two HALO Photonics scanning wind lidars, model StreamLine (SL88) and StreamLine XR (SLXR142), which are run semi-operationally at the University of Innsbruck as part of the Innsbruck Atmospheric Observatory \citep[IAO; ][]{Karl2020}. A recent application of these lidars within IBK is described in \citet{Haid2020}. The SLXR142 is still operating on the rooftop of Innrain~52f and has been performing plan position indicator (PPI) scans at an elevation angle of 70\textdegree~and a range-gate length of 36~m to compute vertical profiles of the horizontal wind for TEAMx-PC22 by applying the velocity-azimuth display (VAD) technique \citep[e.g., ][]{Paeschke2015}. PPI scans took about 80~s each and were repeated continuously. Between 11~August and 02~October~2022, the SL88 lidar was operated next to the Raman lidar and performed vertical stares to measure vertical profiles of the vertical wind component at an interval of 1~s and with a range-gate length of 30~m. Together with the temperature and humidity profiles of the collocated PPL, these high-frequency wind profiles can be used to compute sensible and latent heat flux profiles \citep[e.g., ][]{Behrendt2020,Lareau2020}. Part of the IAO at the University of Innsbruck is a semi-automatic weather station (so-called TAWES), jointly operated by ACINN and the Austrian national weather service GeoSphere Austria (formerly ZAMG). In addition, ACINN continuously operates various instruments on a tower at the rooftop of Innrain~52f to measure turbulence, including turbulent fluxes of various trace gases \citep[see, e.g., ][]{Karl2020,Ward2022}. More recently, these measurements have been extended to the adjacent street canyon.

Radiosondes are operationally released at Innsbruck Airport (WMO station identifier 11120) by the aviation weather service of Austro Control GmbH once per day at 0215~UTC in the summer half-year and at 0315~UTC in the winter half-year. For TEAMx-PC22, ACINN and KIT/IMK-IFU requested additional soundings from Austro Control on 14 selected fair-weather days between 23~August and 24~September 2022 for the calibration and evaluation of the Raman lidar, with one of these days serving as the calibration reference. These additional soundings, 41 in total, were performed typically at about 0800, 1400 and 2000~UTC to achieve an approximate interval of 6 hours between each sounding, starting with the first operational sounding at about 0200~UTC. Some of the radiosondes were released about one hour earlier or later than these target times or could not be released at all due to technical difficulties or other constraints. Data of all soundings were distributed via the Global Telecommunication System (GTS) network and, thus, are publicly available from various radiosonde archives\footnote{For example from \url{https://weather.uwyo.edu/upperair/bufrraob.shtml}}.

\subsection{Sub-target area Kolsass (KOL)\label{sec:KOL}}
Due to easy access to power connection, multiple remote sensing instruments could be deployed for measuring processes on multiple scales within the sub-target area KOL (Fig.~\ref{fig1:target}c). Roughly in the middle of KOL near the village Kolsass a station with a 17-m high turbulence tower\footnote{A full description of the site can be found under \url{https://acinn-data.uibk.ac.at/pages/i-box-kolsass.html}} is operated since 2014 and is part of the Innsbruck box \citep[i-Box; ][]{Rotach2017}. Usually, the tower is labeled CS-VF0 (valley floor, 0\textdegree~slope angle) in \cite{Rotach2017}, however, for near-surface observations of this publication we simply refer to it as KOL in Fig.~\ref{fig4:awsaug} and Fig.~\ref{fig10:awsjun}. For revealing near-surface processes and to connect the upper structure of the MoBL with the surface profile, a DTS array was installed and is described in detail in \cite{Pfister2023_InnDEX22}. The strength of DTS is its spatial resolution of 0.127~m outperforming any point observation \citep[e.g. ][]{Peltola2021, Fritz2021, Pfister2021a}. The technique is well established within atmospheric sciences and can be used in different environments \citep[e.g. ][]{Abdoli2023, Hilland2022, Lapo_2022_LOVE, Karttunen2022, Zeller2021, Schilperoort2020, Shanafield2018}. DTS data are available from 08 to 16~June and from 28~June to 18~July 2022. Temperatures were measured with two channels at 1~Hz with a spatial resolution of 0.127~m. The array was a double-ended configuration but measurements were performed as two single-ended configurations which is different from the manufacturer's provided double-ended mode  \citep{desTombe_2020, Lapo_2022_LOVE}. DTS data were calibrated using the weighted least squares approach described in \cite{desTombe_2020} and implemented in the \textit{dtscalibration} software package \citep{des_tombe_dtscalibration} and all processing was completed using the \textit{pyfocs} software package \citep{lapo_pyfocs}. After calibration a mean bias of $-0.05$~K and root mean squared difference of 0.22~K were determined with the validation water bath.  

Based on the experiences gained during CROSSINN \citep{Adler2021,Babic2021}, Urban Climate Under Change [UC]\textsuperscript{2} \citep{Adler2020,Kiseleva2021,Wittkamp2021}, and High Definition Clouds and Precipitation for Advancing Climate Prediction HD(CP)\textsuperscript{2} \citep{Traeumner2015}, the Institute of Meteorology and Climate Research - Department Troposphere Research (IMK-TRO) of the Karlsruhe Institute of Technology (KIT) combined the KITcube measurement system \citep{Kalthoff2013} at the valley bottom with two additional Doppler wind lidars at the slopes to investigate the outflow of the tributary Weer Valley from 19 May to 20 September 2022. For resolving the MoBL vertically, a Leosphere Windcube WLS200s Doppler wind lidar (WLS200s-115) was installed in the vicinity of the i-Box station. The WLS200s-115 performed a routine loop consisting of a PPI scan at an elevation angle of 70\textdegree{} and a Doppler beam swinging (DBS) scan at the same elevation angle. This scanning strategy resulted in three to five scans of each type within a 10-min period. In this case, the goal was on the one hand to assess the performance of different horizontal wind sampling strategies (DBS vs. VAD) in complex terrain and on the other hand to establish an along-valley transect of lidars together with the other Doppler wind lidars from ACINN (IBK), GeoSphere Austria (IVE), and DWD (IVE). Both types of scans used a custom defined range gate resolution of 50~m and an accumulation time of 1~s.  Vertical wind measurements are taken from the DBS scans as they contain one vertical beam with a measuring time of 4\,s. The PPI/DBS loop ran consistently between 08~June and 17~September~2022, but a data gap of 14 days was caused by a full hard disk in July. In addition, a smaller Windcube v2.1 (WCv2.1) Doppler wind lidar was installed next to the WLS200s-115. Based on DBS scans at an elevation angle of 62\textdegree{} and with a total of 19 range gates spaced linearly apart between 40 and 400~m above valley floor, the WCv2.1 lidar provided horizontal and vertical wind information at the start of every minute from 20~May until 19~September~2022. Lastly, the KITcube RPG FMCW dual frequency 94/35 GHz scanning cloud radar was also deployed within KOL and repeated a scan pattern every 10 minutes which consisted of a PPI scan at an elevation angle of 70\textdegree{} and a period of vertical stares for more than 8~minutes. For the PPI, the antenna was rotated by 5\textdegree~s$^{-1}$ and data were stored at a 1-s interval. The vertical stare was performed with a measurement interval of 10 s. The lowest measurement range was at 100~m, the range resolution was 15~m below a distance of 2000~m and roughly 38~m above.

For investigating the outflow from the Weer Valley and its interaction with the flow in the Inn Valley, two horizontally scanning Leosphere Windcube WLS200s Doppler wind lidars were installed on the northern sidewall near the villages of Fritzens (WSLS200s-159) and Terfens (WLS200s-124). The lidars were located 66 and 64~m above the valley floor with unobstructed line of sights towards the exit region of the Weer Valley. A near identical height above the valley floor was necessary to sample the nocturnal tributary outflow in a purely horizontal plane with PPI scans at 0\textdegree~elevation and applying the coplanar-retrieval method, also known as the dual-Doppler method, to radial velocity measurements \citep{Traeumner2015,Haid2020,Haid2022,Adler2021,Babic2021,Babic2023}. Those PPI scans were started at the beginning of even minutes, resulting in a duration of 106~s, an accumulation time of 1~s, and overlapping 50-m range gates spaced apart every 25~m. This scanning strategy of just PPI scans lasted between 29~June and 31~August 2022. 

For testing and intercomparison, two new short-range Doppler wind lidars (Wind Ranger~200), manufactured by METEK and owned by ACINN and GeoSphere Austria, were operated next to the KITcube Doppler wind lidars at the i-Box station between 09~June and 07~August 2022. They are based on the frequency modulated continuous wave (FMCW) method and provide vertical profiles of the 3D wind vector in the lowest 200~m above ground level (AGL), with the lowest height at 10~m AGL, based on PPI scans at a fixed zenith angle of 10\textdegree. From now on, these lidars will be called WR21 (ACINN) and WR19 (GeoSphere). In addition to instrument testing, the aim of their application was to fill the data gap between the highest measuring height of the Kolsass i-Box flux tower and the lowest measuring height of the KITcube lidars. When comparing the two Wind Ranger lidars, an error was found in the real-time data processing software, which meanwhile has been fixed by the manufacturer. However, for the TEAMx-PC22 period, this error excludes the usability of the collected data for research purposes before the error was fixed.

\subsection{Sub-target area Nafingalm (NAF) \label{sec:NAF}}
The sub-target area NAF (Fig.~\ref{fig1:target}d) was chosen to investigate small-scale thermodynamic processes in a remote Alpine location by combining a network of surface observations with a small fleet of quadrotor UAS and to test these UAS over complex terrain. The UAS are part of the ESTABLIS-UAS project\footnote{\url{https://establis-uas.eu/}} of the Deutsches Zentrum für Luft- und Raumfahrt e.V. (DLR). In order to investigate how three-dimensional turbulence manifests, spatially distributed sensors are required ideally up to the boundary layer height, but also at the ground. Accordingly, ACINN supported DLR's efforts with further ground-based observations.  The setup with ground-based observations and UAS should make it possible to validate the representation of turbulence within NWP models as well as to investigate the scale interaction between thermally driven winds and mesoscale as well as synoptic forcings. The UAS fleet was already part of the Field Experiment on Submesoscale Spatio-Temporal Variability in Lindenberg \citep[FESSTVaL, ][]{FESSTVaL} within which the data was calibrated and validated \citep{Wetz2021} and revealed coherence in flat but heterogeneous terrain \citep{Wetz2023}. Within TEAMx-PC22 the UAS fleet is now tested over heterogeneous and complex terrain. 

From~20 to 28~June 2022, three UAS were operated simultaneously in different configurations and flight strategies. Hover flights, vertical profiles up to 120~m AGL, and flights crossing the valley horizontally at height levels 10~m and 40~m above the valley floor were the main flight strategies that were pursued. Some of the patterns were combined by up to three UAS flying simultaneously, spread along the valley axis or aligned on a cross-valley axis. Data of the ascending vertical profiles and time series of fixed-point measurements during hover flights can be found in \cite{wildmann:23}. The average flight duration was about twelve minutes and was usually conducted multiple times during the day. Additionally, ground-based instrumentation was deployed between 15~June and 12~September 2022 to support the UAS measurements and to get a better statistical understanding of the processes in the valley. An AWS was operated north of the lake and in its vicinity a three-dimensional ultrasonic anemometer was measuring turbulence during the period of UAS activities. Four stations equipped with multi-level temperature and humidity sensors were installed north and south of the lake, on the eastern valley slope, and on the mountain crest. Lake water temperature was measured on a tethered buoy at two depths.

\subsection{Sub-target area Inn Valley Exit (IVE) \label{sec:IVE}}
In the sub-target area IVE (Fig.~\ref{fig1:target}e), the Inn Valley exit-jet was investigated. Therefore, the German Meteorological Service (DWD) deployed a Doppler wind lidar and a small surface station network to extend previous findings and to provide a reference dataset for testing the PALM modelling system for mountain boundary-layer applications \citep{Maronga2020}. The Austrian national weather service GeoSphere Austria added ground-based remote sensing measurements in order to observe the boundary layer structure and the lower part of the valley flow in the upstream section of the IVE.

At Kufstein in the southernmost part of IVE, a ceilometer (Vaisala CL51) was operated next to a TAWES weather station, both of which are part of the permanent observation network of GeoSphere Austria. The CL51 provided vertical profiles of attenuated backscatter and derived aerosol layer heights at a time interval of 36~s and a range-gate length of 10~m. From 17~August to 11~November 2022, the site was equipped with a short-range Doppler wind lidar (Wind Ranger 200, METEK), hereinafter referred to as WR13. The before mentioned software error was fixed for these measurements. Similar to WR19 und WR21 at KOL, WR13 provided vertical profiles of the 3D wind vector in the lowest 200~m AGL. 

The main field site for the exit-jet study was located at the sewage works in Brannenburg (BRA) where an AWS was complemented by a Doppler wind lidar (Halo Photonics StreamLine XR, hereinafter referred to as SLXR172) to continuously capture the vertical profile of the three-dimensional wind vector up to 860~m AGL  \citep{Leinweber2023}. Radial winds were captured with PPI scans in continuous scanning mode (CSM) at an elevation angle of 35\textdegree. Azimuth angle intervals of the CSM data sampling was about 1.1\textdegree. During the period from 27~July to 12~August~2022 the PPI CSM data sampling was replaced by PPI scans in step-stare mode at an elevation angle of 35\textdegree and with azimuth steps of 15\textdegree, complemented by step-stare range-height indicator (RHI) scans pointing into the Inn Valley. The RHI scans covered elevation angles from 3\textdegree~to 51\textdegree~and an azimuth range of 10\textdegree~(i.e., from 150\textdegree~to 160\textdegree).

During the night from 18 to 19~July~2022, vertical soundings with radiosondes and an instrumented drone were performed. They consisted of six drone ascents and nine radiosonde releases at the Brannenburg site  \citep{Paunovic2023b}. Radiosondes started on 18~July at 1203~UTC, subsequent ascents were conducted every three hours from 18~July at 1345~UTC until 0745~UTC 19~July, and followed by the final release on 19~July at 0828~UTC. Parallel to the radiosondes, instrumented drone profiles up to 120~m~AGL were conducted at 1245~UTC 18~July and continued with 3-hourly ascents between 1345~UTC 18~July and 0145~UTC 19~July 2022. 

Further, a network of four surface stations was set up in the IVE and is operated permanently until the end of the TOC \citep{Paunovic2023a}.  Three stations are located along the Inn Valley floor in Flintsbach (FLI), Oberaudorf, and Kiefersfelden to capture along-valley variations of air temperature, humidity, wind speed, and pressure associated with nocturnal downvalley winds. The surface stations at BRA and Kiefersfelden are about 10~km apart. A pronounced valley constriction between the stations Oberaudorf and FLI is thought to contribute to the formation and intensity of the Inn Valley exit-jet. To continuously monitor the approximate bulk thermal stratification and winds above the valley, an additional weather station is located at the Hohe Asten, a mountain top on the western Inn Valley side wall near the valley constriction.

\section{Exemplary case studies}
The instrument setup of the TEAMx-PC22 made it possible to resolve a range of scales both horizontally and vertically. Three different events were chosen to demonstrate the strength of this multi-scale dataset to capture different atmospheric processes in the IVTA. Since not all instruments were operating continuously during TEAMx-PC22, a good compromise between data availability and the occurrence of interesting processes had to be made. The first event (24~August 2022) is characterized by weak synoptic forcing and the focus is on processes in the main valley and at the valley exit region (Sec.~\ref{case1}). The second event (23~June 2022) features moderate synoptic forcing with south-foehn influence and the focus is on processes in tributary valleys, specifically the Weer Valley at NAF (Sec.~\ref{case2}). For both events, the synoptic scale condition is illustrated with ERA5 reanalyses \citep{Hersbach2020} and the mesoscale situation is described based on operational COSMO-1 analyses and forecasts (3 UTC run) provided by Meteo Swiss. The third event (17~July 2022) was chosen to test and demonstrate the capability of DTS and cloud radar measurements, that were not available for the other two events (Sec.~\ref{sec:further_measurements_tested}).

\subsection{Valley winds during weak synoptic forcing}\label{case1}
\subsubsection{Synoptic and mesoscale overview}\label{sec:synop1}
\begin{figure}
    \centering
    \includegraphics[width = \textwidth]{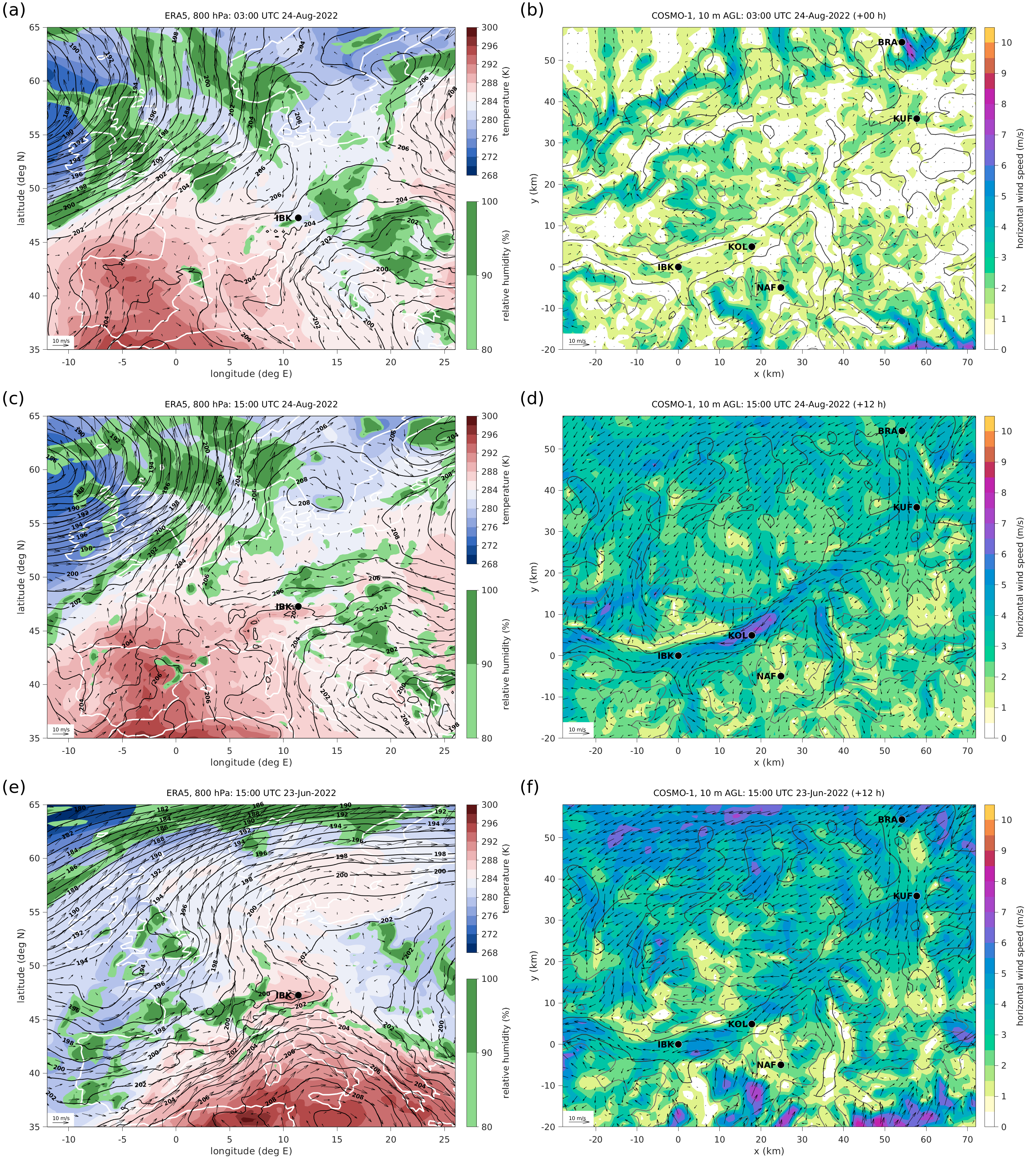}
    \caption{Synoptic and mesoscale condition at (a)-(b) 0300~UTC, (c)-(d) 1500~UTC 24~August 2022 and (e)-(f) 1500~UTC 23~June 2022: ERA5 temperature and relative humidity (color contours) and geopotential height (black contours; 1 gdm increments) and the horizontal wind field at 800 hPa over Europe in (a), (c) and (e). COSMO-1 horizontal wind field at 10~m AGL (color contours for wind speed) in the Inn Valley target area in (b), (d) and (f). Horizontal wind fields shown by wind vectors (reference vector in lower left corner). Markers and labels show the location of Innsbruck (IBK), Kolsass (KOL), Nafingalm (NAF), Kufstein (KUF), and Brannenburg (BRA). The COSMO-1 analysis is shown in (b) and the 12-hour forecast in (d) and (f).}
\label{fig3:synop}
\end{figure}

The first case study on 24~August 2022 is characterized by weak synoptic forcing (Fig.~\ref{fig3:synop}a,c for 800-hPa ERA5 analysis and Fig.~\ref{fig3:synop}b,d for COSMO-1 surface winds). At 0300~UTC a weak high-pressure system over central Europe is bounded by two low pressure systems located over the Atlantic west of the British Isles and over the Eastern Mediterranean  (Fig.~\ref{fig3:synop}a). The Alps are influenced by weak northeasterly flow around the high. The low-level air mass is colder and moister over the northeastern Alpine foreland compared to the south of the Alps. As the large-scale flow becomes more easterly during the course of the day, this moisture is advected closer to the entrance of the Inn Valley (cf. 0300~UTC and 1500~UTC in Fig.~\ref{fig3:synop}a,c at location IBK).

As a result of the weak synoptic forcing, a well-developed valley wind system establishes in the IVTA over the entire period. During the night, the Inn Valley is governed by weak thermally-driven downvalley winds (Fig.~\ref{fig3:synop}b) which can be identified in the COSMO-1 analysis in all sub-target areas (see station labels in Fig.~\ref{fig3:synop}b). The strongest winds occur near BRA where the flow forms a valley exit-jet. During daytime, thermally-driven upvalley winds develop in the Inn Valley (Fig.~\ref{fig3:synop}d). The low-level pressure gradient induced by the large-scale temperature contrast across the Alps may be a reinforcing factor for the inflow from the Alpine foreland. The strongest winds in the Inn Valley occur near KOL (cf. Fig.~\ref{fig3:synop}d). The locations of the strongest winds during daytime and nighttime are in agreement with previous studies \citep[e.g., Fig. 2 in][]{Zangl2004}.

\subsubsection{Valley winds in the IVTA \label{sec:valley_winds_IVTA}}
\begin{figure}
\centering
   \includegraphics[width=\textwidth]{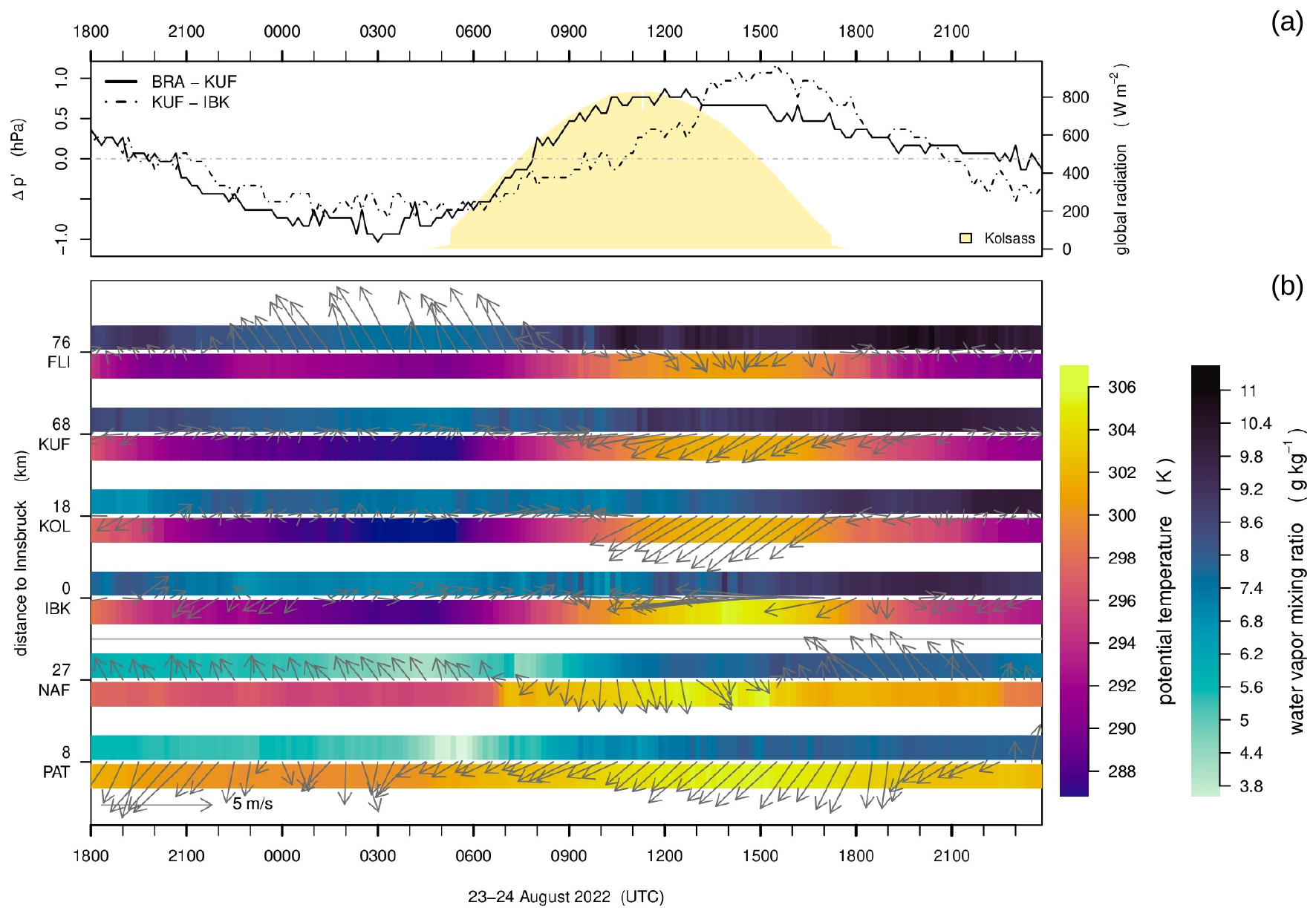}
   \caption{Near-surface observations along the Inn Valley target area at the stations Innsbruck University (IBK), Kolsass (KOL), Kufstein (KUF), and Flintsbach (FLI) from 1800~UTC 23~August 2022   to  0000~UTC 25~August 2022. Additionally, the stations at the mountain peak Patscherkofel (PAT) and in the Weer Valley at Nafingalm (NAF) south of the Inn Valley are shown. (a) Time series of global radiation at Kolsass shown as pale yellow area and the horizontal pressure gradient $\Delta p'$ in hPa between near-surface stations shown as lines ($p'$ is the deviation from the temporal mean over the shown period at each location). (b) Stations were grouped as elevated and valley stations, separated by a horizontal grey line. For each station the time series of the potential temperature in K and water vapor mixing ratio in g~kg$^{-1}$ are shown in colours (cf. colorbars on the right), as well as the wind speed and direction as arrows.}
   \label{fig4:awsaug}
\end{figure}

Due to the synoptically undisturbed conditions all sub-target areas exhibit a clear diurnal cycle in temperature and wind, with downvalley (upvalley) flow during the night (day). During daytime only shallow cumuli form (not shown). These conditions in the IVTA permit unperturbed radiative cooling/heating (see, e.g., global radiation at KOL in Fig.~\ref{fig4:awsaug}a).  Hence, not only the flow in the main valley, but also in the tributary Weer Valley are locally forced by radiation, meaning that due to differential boundary-layer cooling/heating within the valleys an along-valley pressure gradient forms (Fig.~\ref{fig4:awsaug}) which drives the associated along-valley winds. 

\begin{figure}
    \centering
    \includegraphics[width = 0.98\textwidth]{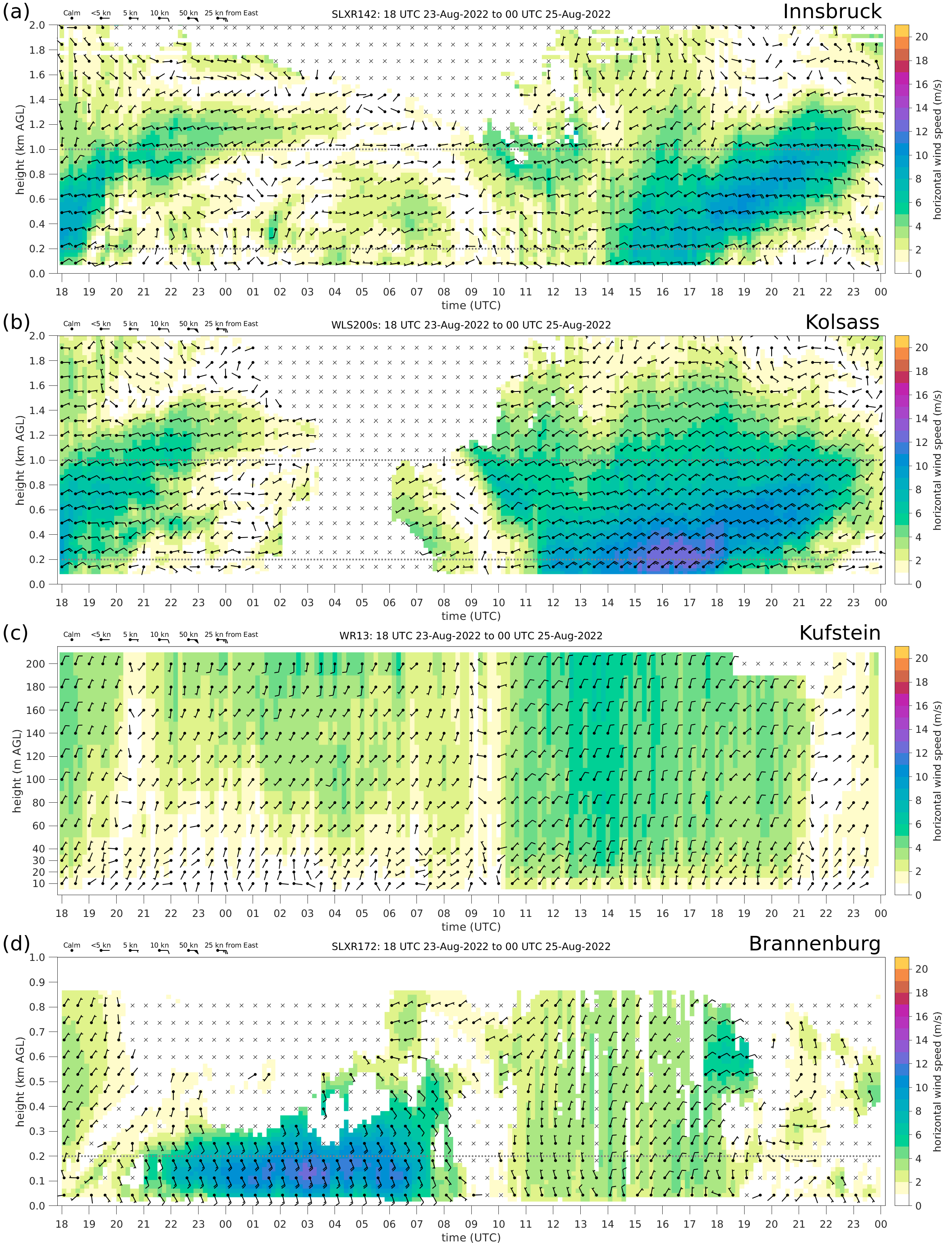}
    \caption{Time-height diagram of 10-min average horizontal wind speed (color shading) and horizontal wind direction (wind barbs) measured between 1800~UTC 23~August and 0000~UTC 25~August 2022 by the Doppler wind lidars (a) SLXR142 at Innsbruck, (b) WLS200s at Kolsass, (c) WR13 at Kufstein, and (d) SLXR172 at Brannenburg, hence, within the sub-target area IBK, KOL, and the last two within IVE. Note the different upper limits of the four vertical axes as a result of the different maximum range of the four lidars. For easier comparison, the two lower top heights, 1000~m AMSL and 200~m AMSL, are indicated with dotted lines where appropriate. Crosses mark missing data.}
\label{fig5:adwl}
\end{figure}

Within the Inn Valley, the morning transition to an upvalley flow happens almost simultaneously between 0900 and 1000~UTC in IBK, KOL, and IVE (Fig.~\ref{fig5:adwl}). Note that the orientation of the valley axis, and thus the wind direction of along-valley winds, are location-dependent (cf. Fig.~\ref{fig1:target}a). While mean wind speed increases at KOL to above 10 m~s$^{-1}$ soon after the wind reversal, it takes several hours longer to reach a comparable magnitude at Innsbruck (Fig.~\ref{fig5:adwl}a,b). The strongest observed upvalley winds of about 12~to 14~m~s$^{-1}$ occur in the afternoon at KOL (Fig.~\ref{fig5:adwl}b). 

Also worth mentioning is the moisture transport by upvalley winds during daytime from the Alpine foreland towards and along the Inn Valley. This moisture advection can be seen from the progressive increase in moisture from western to eastern stations (from FLI to IBK in Fig.~\ref{fig4:awsaug}), starting with the reversal from downvalley to upvalley flow at the valley exit. This along-valley moisture transport is in agreement with lidar and radiosonde measurements. The upvalley flow remarkably increases around 1400~UTC (Fig.~\ref{fig6:libk}a) and  moisture starts to increase (Fig.~\ref{fig6:libk}d and Fig.~\ref{fig7:raso}b). At the same time near-surface turbulence decreases (Fig.~\ref{fig6:libk}b), suggesting that the moisture increase is not caused by vertical turbulent transport, but rather by horizontal advection.

In the evening roughly at 1800~UTC a reversal to downvalley winds occurs when due to the formation of a stable layer the upvalley flow detaches from the surface as observed in IBK (Fig.~\ref{fig6:libk}a). The weak westerly downvalley flow prevails until 0830~UTC (not shown). Comparing the time of the evening transition within the Inn Valley (Fig.~\ref{fig5:adwl}), it is at least two hours earlier at Innsbruck (IBK) and Brannenburg (IVE) than at Kolsass (KOL) and Kufstein (IVE). The reason for this along-valley heterogeneity in the evening transition is counterintuitive at first glance and needs to be explored in the future. 

A striking feature during nighttime are strong southeasterly (downvalley) winds at the exit of the Inn Valley (see FLI in Fig.~\ref{fig4:awsaug}). Its onset at around 2100~UTC causes transient warming, probably due to turbulent mixing while stronger cooling occurs further upstream at Kufstein where winds are weaker (cf. FLI and KUF in Fig.~\ref{fig4:awsaug}). Kufstein is located only 20~km upstream of Brannenburg and about 10~km of the narrowest section of the valley exit region. Lidar measurements detect the strongest winds at about 0300~UTC at Brannenburg at the exit of the Inn Valley (Fig.~\ref{fig5:adwl}d). Mean winds exceed about 12 m~s$^{-1}$ at 150~m AGL and form the well-known low-level exit-jet \citep[e.g., ][]{Pamperin1985,Zangl2004}. At the same height at Kufstein the nighttime wind speed is only about 4 m~s$^{-1}$. Hence, winds accelerate downstream of the narrowest section. This acceleration can be interpreted as a transition from subcritical to supercritical hydraulic flow \citep[e.g., ][]{Zangl2004}.

\subsubsection{The convective boundary layer and subsidence of dry air in the Inn Valley \label{sec:CBLs}}
\begin{figure}
    \centering
    \includegraphics[width = 0.9\textwidth]{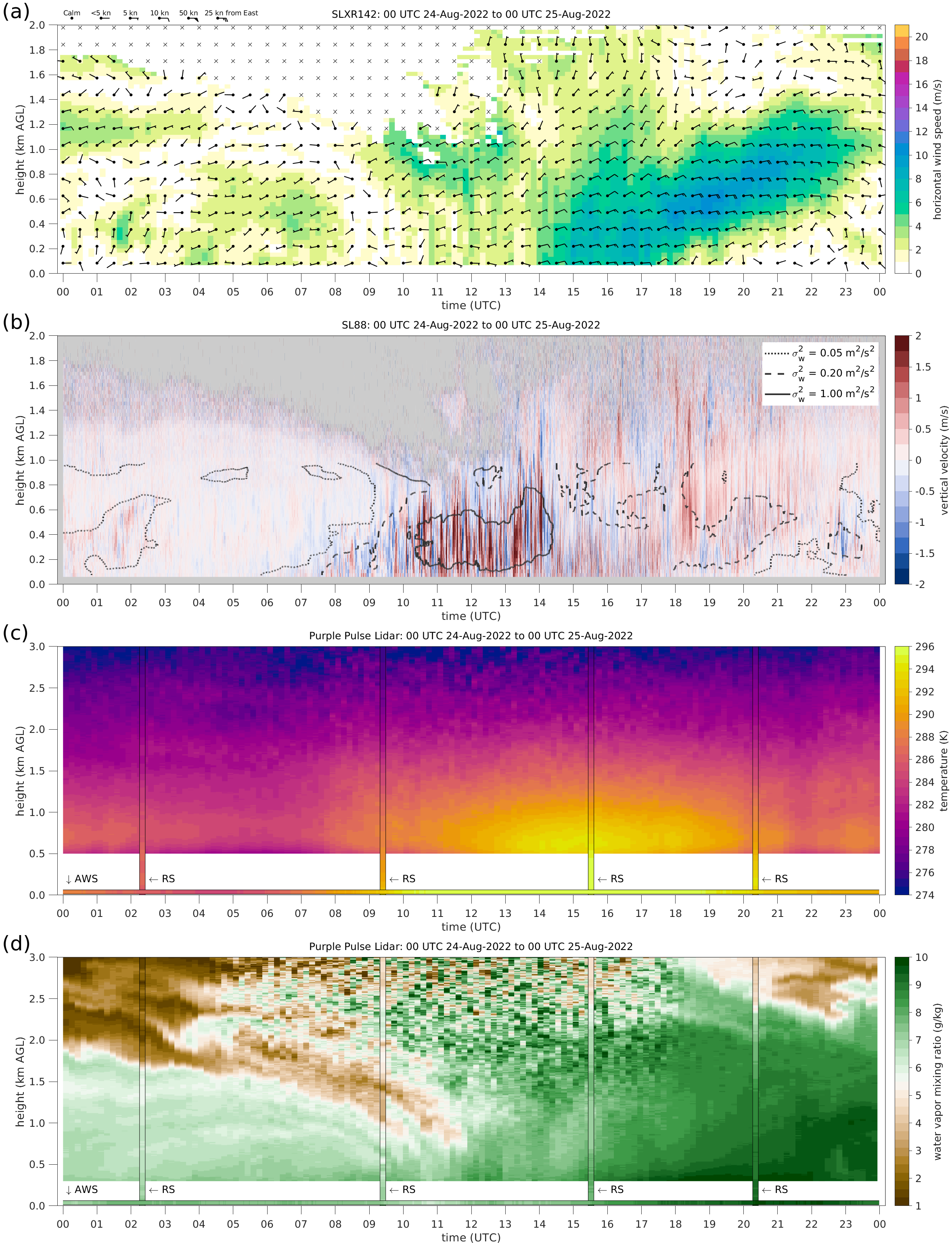}
    \caption{Time-height diagram illustrating the evolution and structure of the valley atmosphere at IBK over 24 hours on 24~August 2022 as observed by different remote-sensing and in-situ instruments: (a) 10-min average horizontal winds (color shading for speed and barbs for horizontal direction) measured with the Doppler wind lidar SLXR142. (b) Vertical wind component (1-s resolution, color contours) measured with the Doppler wind lidar SL88 and associated vertical velocity variance (black contours for 0.05, 0.2 and 1~m$^2$~s$^{-2}$; see legend) calculated as the moving variance over a sliding time window of 1 hour. (c) Temperature and (d) water vapor mixing ratio measured with the Raman lidar PPL (10-minute averaging, color shading), four radiosondes launched at Innsbruck Airport (vertical color bars), and the weather station TAWES at the University of Innsbruck. Note that the vertical axis goes up to 2 km AGL in (a) and (b) and to 3 km AGL in (c) and (d). Noisy daytime Raman lidar data above about 1.5 km were not filtered to retain some of the coarser structure.}
\label{fig6:libk}
\end{figure}

\begin{figure}
    \centering
    \includegraphics[width = \textwidth]{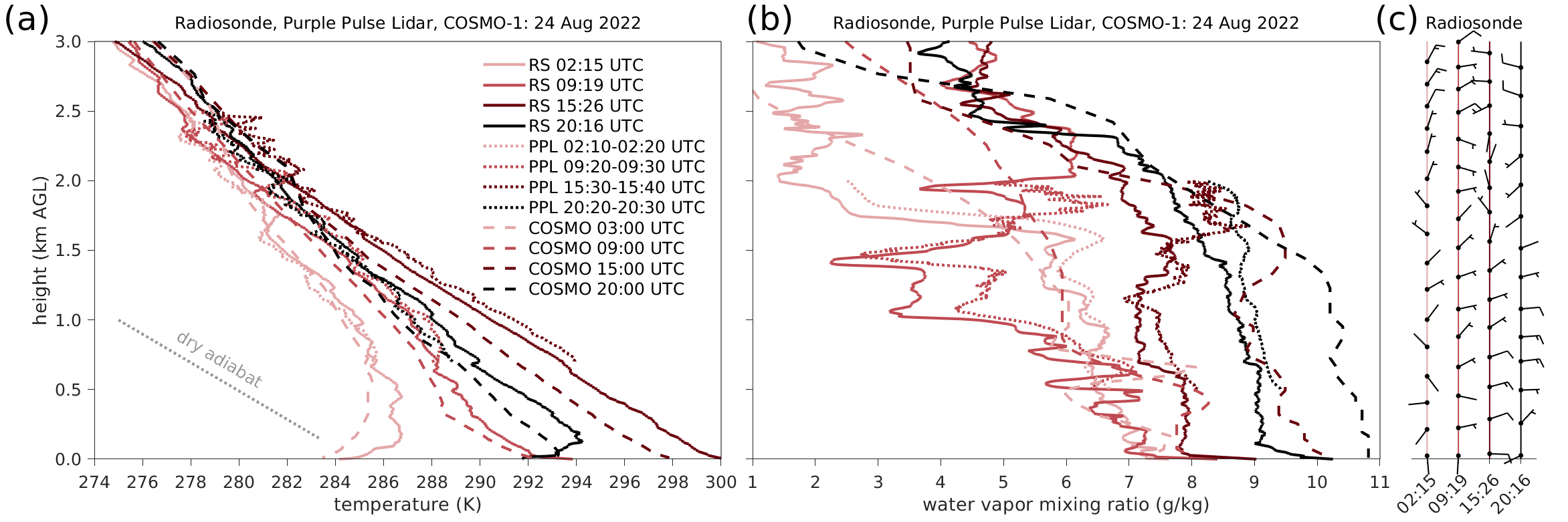}
    \caption{Vertical profiles of (a) temperature and (b) water vapor mixing ratio at Innsbruck on 24~August 2022 at four different times measured by the radiosonde (RS, solid lines) and the Raman lidar (PPL, dotted lines) and predicted by the operational COSMO-1 forecast of Meteo Swiss initialized at 0300~UTC (dashed lines). Different times are indicated by different colors. The legend indicates the launch time of the radiosonde, the averaging period for the Raman lidar data and the time of the instantaneous model output. Wind barbs in (c) represent the horizontal wind measured by the radiosondes (see legend for barbs in Fig.~\ref{fig6:libk}a).}
\label{fig7:raso}
\end{figure}

\begin{figure}
    \centering
    \includegraphics[width = \textwidth]{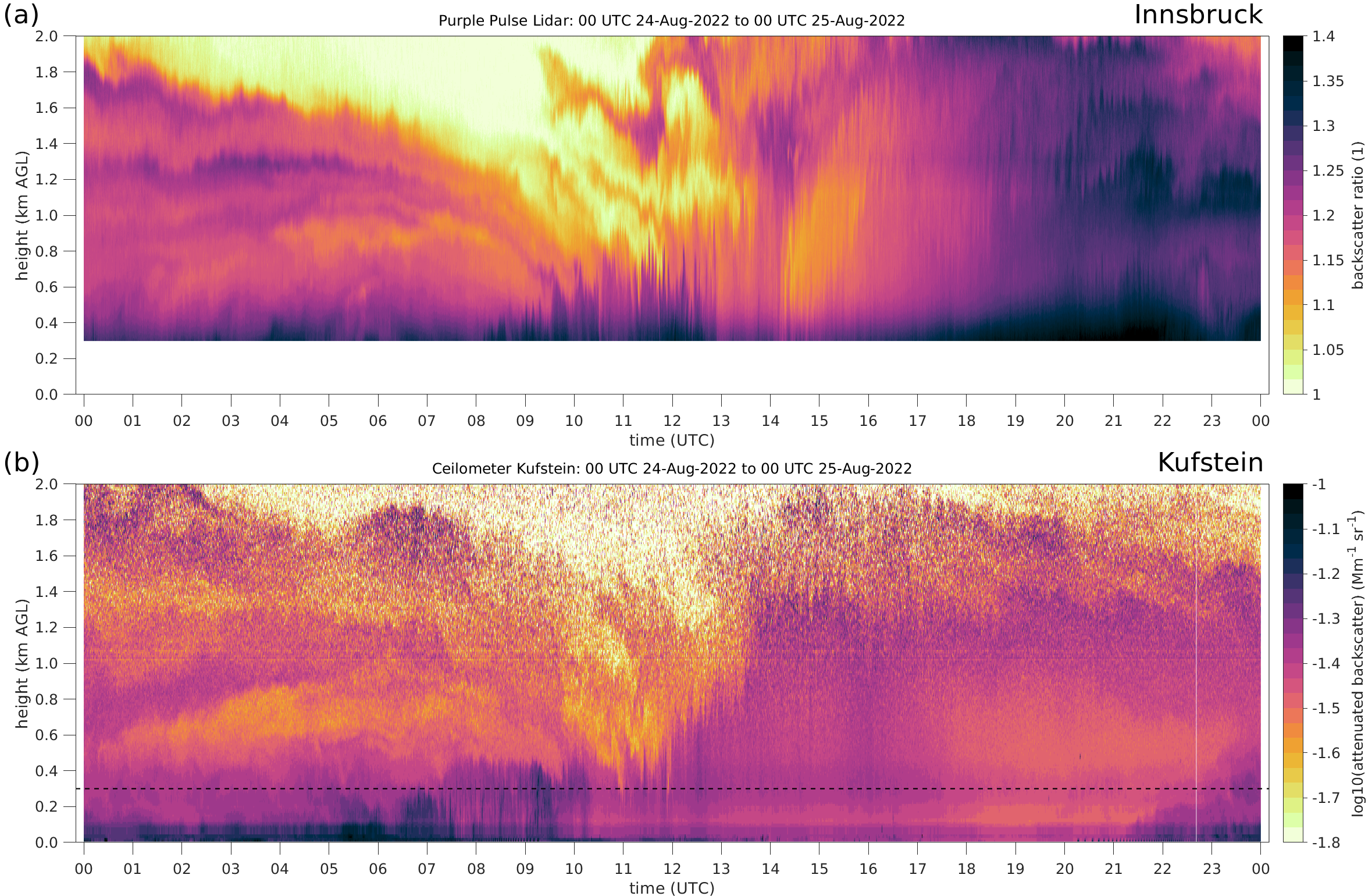}
    \caption{Time-height diagram of the aerosol backscatter measured over 24 hours on 24~August 2022 by (a) the Raman lidar at Innsbruck and (b) the ceilometer at Kufstein. Note that the profiles in (a) start at 300~m AGL. This reference height is indicated in (b) as a dashed line.}
\label{fig8:ceil}
\end{figure}

After sunrise at about 0500~UTC, near the surface a convective boundary layer (CBL) starts to develop in the sub-target area IBK. Note that the CBL does not constitute the entire MoBL, but only its lower turbulent part. The CBL is indicated by the vertical velocity variance exceeding 0.2~m$^2$~s$^{-2}$  (Fig.~\ref{fig6:libk}b) which is caused by consecutive up- and downdrafts within thermal plumes. Until about 0900~UTC the CBL has grown to a depth of about 400~m, which agrees with the depth of the dry adiabatic layer in the observed temperature sounding (Fig.~\ref{fig7:raso}a). Afterwards the CBL grows beyond 1~km AGL, but the lack of lidar backscatter signal (see also Fig.~\ref{fig8:ceil}a) does not allow a clear detection of the CBL top due to subsidence of dry air (Fig.~\ref{fig6:libk}d). 

Signatures of thermal plumes in the CBL manifest as vertical stripes of variable backscatter intensities measured by the Raman lidar and thus variable aerosol loads (Fig.~\ref{fig8:ceil}a). These stripes correlate qualitatively well with up- and downdrafts detected by the Doppler wind lidar (Fig.~\ref{fig6:libk}b). Such thermal plume signatures can also be qualitatively observed by the ceilometer at Kufstein (Fig.~\ref{fig8:ceil}b) indicating the CBL depth. But there the CBL at 1200~UTC is shallower (500~m AGL) than at Innsbruck (800~m AGL). During the phase of fully developed upvalley winds in the afternoon and the associated vertical mixing over a deeper layer (Fig.~\ref{fig6:libk}a-b), the backscatter profiles at both sites lack coherent structures and are therefore less suitable for detecting the boundary layer structure. 

Subsidence of a layer of dry air with a low aerosol load above IBK is well captured by the Raman lidar (Fig.~\ref{fig6:libk}d) and the two early radiosonde profiles (Fig.~\ref{fig7:raso}b). The bottom of this layer is located at about 2~km AGL at sunrise and subsides to below 1~km AGL at noon. This subsidence of dry air most likely happened also at other sites along the IVTA as it is also observed by the ceilometer at Kufstein (Fig.~\ref{fig8:ceil}b). The mechanism causing the subsidence of dry air is unclear, but might be the compensation for the thermally driven upslope winds or might be terrain-induced by the northeasterly cross-Alpine flow. The subsidence is weak, only about $2$~cm~s$^{-1}$ estimated from the evolution of the backscatter and moisture profiles. Nevertheless, it does not limit the growth of the CBL in the afternoon, since the observed dry adiabatic layer reaches beyond 1.5~km AGL at about 1500~UTC (Fig.~\ref{fig7:raso}a).

\subsubsection{Cold-air pools, outflow from tributaries, and scale interactions \label{sec:CAP_and_Tributaries}}
In the late afternoon on 24~August 2022 a stable boundary layer forms in IBK  (Fig.~\ref{fig6:libk}c; see also temperature inversion at 2000~UTC in~\ref{fig7:raso}a). The continuous growth of this nighttime cold-air pool (CAP) results in a continuous lifting of the upvalley flow and a reduction of its depth until about midnight (Fig.~\ref{fig6:libk}a). At IBK the CAP is about 200 m deep at 2000~UTC (Fig.~\ref{fig7:raso}a) and characterized by weak turbulent mixing (Fig.~\ref{fig6:libk}b). Close to midnight, a weak southerly flow establishes in the upper part of the CAP (e.g., 2330 UTC in Fig.~\ref{fig6:libk}a). These southerlies are even more pronounced in the previous night (e.g., 0200~UTC in Fig.~\ref{fig6:libk}a and represent the outflow from the Wipp Valley, a tributary south of IBK (cf. Fig.~\ref{fig1:target}a). The outflow rides over the CAP in the Inn Valley and causes moderate turbulent mixing at the top of the CAP, as depicted by vertical velocity variances exceeding 0.05 m$^2$~s$^{-2}$ (see 0200~UTC in Fig.~\ref{fig6:libk}b).

\begin{figure}
    \centering
    \includegraphics[width = 0.8\textwidth]{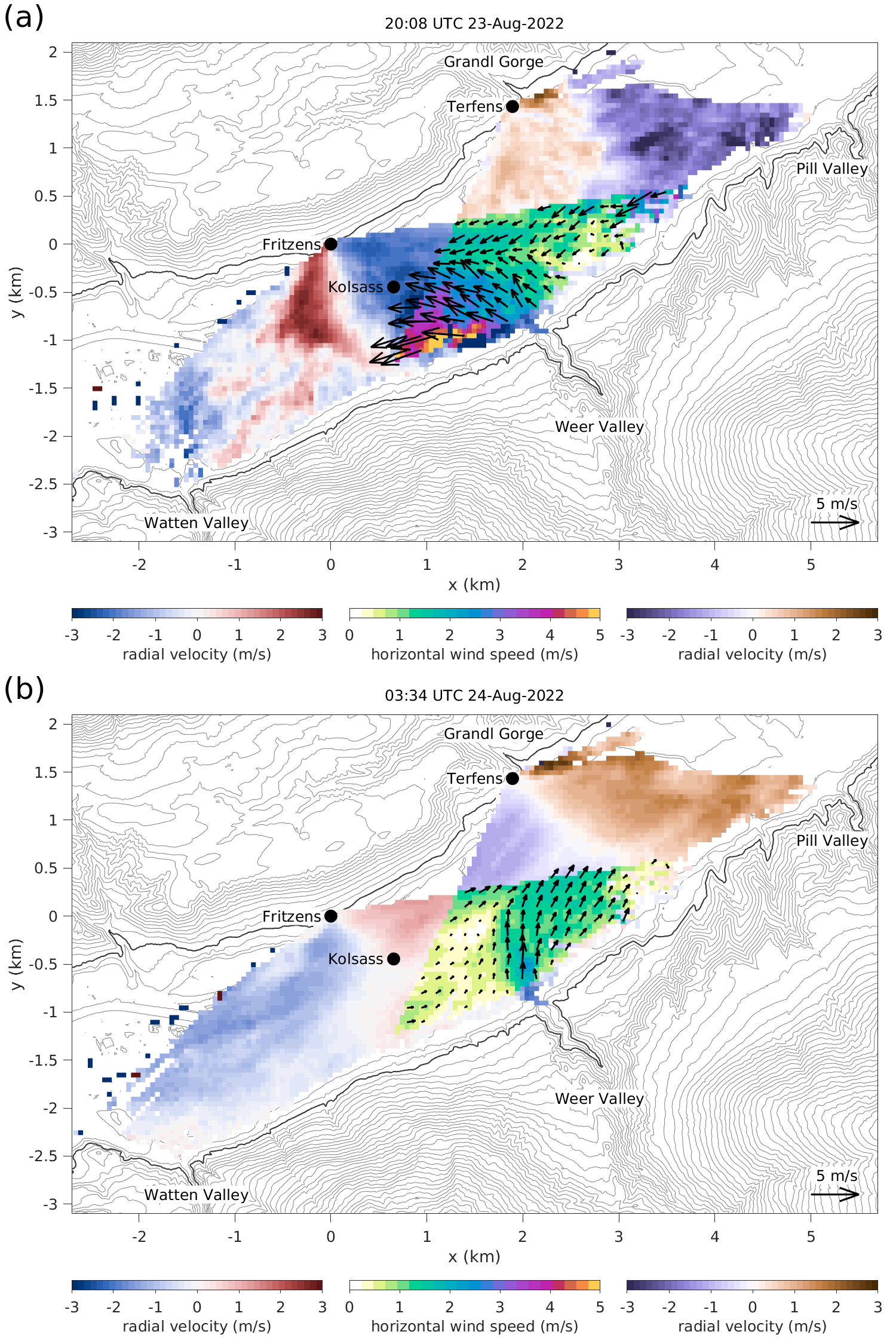}
    \caption{Horizontal wind field at about 65~m above the valley floor at (a) 2008~UTC 23~August and (b) 0334~UTC 24~August 2022 measured by the two WLS200s Doppler wind lidars located at Fritzens and Terfens on the northern slope of the Inn Valley. Radial velocities are negative (positive) for flow towards (away from) the respective lidar and are color shaded with two different color palettes (blue-red for Fritzens  and purple-brown for Terfens). Coplanar retrievals of horizontal wind speed (color shading, third color palette) and direction (arrows) are shown in the region where the two lidar scanning planes overlap and where the dual-Doppler analysis can be applied. Terrain elevation is illustrated by thin gray contour lines with 20~m increments and the approximate height of the lidar sites (610~m AMSL) is highlighted by a thick black contour line. Major tributary valleys are labeled.}
\label{fig9:copl}
\end{figure}

In the evening of 23~August 2022 a similar interaction can be observed for the Weer Valley in the sub-target area KOL where a more detailed analysis is possible based on the dual-Doppler method for retrieving the horizontal wind speed and direction at about 65~m above the Inn Valley floor (cf. Section~\ref{sec:KOL}). The outflow from the Weer Valley around 2000~UTC is deflected westward by the still ongoing upvalley flow in the Inn Valley (Fig.~\ref{fig9:copl}a). The outflow is too shallow to be detected with the Doppler wind lidar at KOL, whose profile starts at about 100~m AGL (Fig.~\ref{fig5:adwl}b). Presumably the colder tributary outflow penetrates underneath the upvalley winds and causes a complex transient low-level flow pattern at the Weer Valley while due to the transient nature and weak flow the near-surface stations in KOL could not detect the cold air flowing out of the Weer Valley. In this early stage the tributary outflow is strongest and partly exceeds 5~m~s$^{-1}$. A few hours later, around 2200~UTC, the upvalley flow in the main valley has ceased (Fig.~\ref{fig5:adwl}b) and the tributary outflow penetrates the valley atmosphere without mean deflection but with considerable meandering and oscillation \citep[not shown; see animations of][]{Babic2022}. With the onset of downvalley winds in the main valley in the second half of the night, the tributary outflow is again deflected, but now eastward in downvalley direction (e.g., 0334~UTC 24~August in Fig.~\ref{fig9:copl}b). The tributary outflow is now weaker and hardly exceeds 2~m~s$^{-1}$.

\subsubsection{Instrument comparison and model validation \label{sec:model_comparison_case1}}
Since the Raman lidar is a new instrument within the IVTA, it was compared to nearby radiosonde measurements. Due to the complex terrain and not probing the same volume an instrument validation by direct intercomparison can become rather challenging \citep{Vogelmann2015}. Significant differences were observed in particular during convective phases. In general, the agreement is good as temperature and water vapor profiles typically differ less than 1~K and 0.5~g~kg$^{-1}$, respectively  (Fig.~\ref{fig6:libk}c-d, Fig.~\ref{fig7:raso}). While radiosondes have a better spatial resolution and no blind zone at low elevations, the Raman lidar measures continuously and can fill gaps between radiosonde launches above the lidar blind zone. Accordingly, the combination of both systems greatly facilitates the detection and interpretation of boundary layer processes in complex terrain such as the previously mentioned horizontal and vertical moisture advection. 

Comparing the measurements with the COSMO-1 forecast, the overall evolution of the valley atmosphere is represented reasonably well in the model (Fig.~\ref{fig7:raso}). Moderate northeasterly winds of less than 5~m~s$^{-1}$ at the mountain peak station Patscherkofel (2251~m AMSL; see Fig.~\ref{fig1:target}a) are in line with the large-scale flow (see Sec.~\ref{sec:synop1}). They prevail all day and advect moist air from the northeastern Alpine foreland (see moisture increase at PAT in Fig.~\ref{fig4:awsaug}). But the model underestimates temperatures in the stable and convective boundary layer by 1~to 2~K (Fig.~\ref{fig7:raso}a). Moreover, the subsidence and associated import of dry air  (Fig.~\ref{fig7:raso}b) is not captured by COSMO-1. Last but not least, the moisture increase in the afternoon due to advection by the upvalley flow is too strong in the model and results in moisture being overestimated by up to 2~g~kg$^{-1}$, which is equivalent to an overestimation of up to 25\%.

The strongest observed upvalley winds of about 12~to 14~m~s$^{-1}$ occur in the afternoon at KOL (Fig.~\ref{fig5:adwl}b). This location of the strongest upvalley flow is in agreement with the COSMO-1 forecast (Fig.~\ref{fig3:synop}d) and earlier model results \citep[Fig. 2b and 3a in ][]{Zangl2004}. The increasing wind speed from the IVE to KOL (cf. Fig.~\ref{fig5:adwl}b-d) may indicate that the vertically integrated mass flux along the Inn Valley is increasing too, which would require horizontal mass import through northern tributaries \citep[e.g., the Achen Valley as suggested by][]{Zangl2004} or vertical import through subsidence. However, a mass budget analysis based on model data would be required for a quantitative answer \citep[e.g., ][]{Deidda2023}.

\subsection{Valley winds perturbed by foehn}\label{case2}
\subsubsection{Synoptic and mesoscale overview}\label{sec:synop2}
During the second event on 23~June 2022 the Alps are located between a trough over the Atlantic and a ridge over Eastern Europe causing prevailing large-scale winds near the Alpine crest level from the southwest at 1500~UTC (Fig.~\ref{fig3:synop}e,f).  A deformation of the isohypses over the Alps forms the so-called foehn nose\footnote{\url{https://glossary.ametsoc.org/wiki/Foehn_nose}} and an associated cross-Alpine pressure gradient (near IBK in Fig.~\ref{fig3:synop}e). Earlier in the morning at 0300~UTC (not shown) the ridge axis is located further west over Central Europe and the associated winds over the Alps are from the northwest, which can also be observed at PAT (Fig.~\ref{fig10:awsjun}b).

In the Inn Valley and its tributaries, thermally driven upvalley winds develop during the course of the day (Fig. ~\ref{fig3:synop}f). However, in some regions they are weaker than during the previous case (cf. Fig.~\ref{fig3:synop}d,f). For example, at 1500~UTC, COSMO-1 predicts no upvalley winds in the Wipp Valley (south of IBK in Fig. ~\ref{fig3:synop}f) and in the southern part of the Weer Valley at NAF (cf. Fig. ~\ref{fig3:synop}f). The reason is that the southerly cross-Alpine flow near crest level (see strongest winds in Fig.~\ref{fig3:synop}f) counteracts the development of northerly upvalley winds in some of the southern tributaries and leads to weak south foehn in the afternoon in some places (not shown). 

\subsubsection{Valley winds and foehn in the Inn Valley}
\begin{figure}
    \centering
    \includegraphics[width=\textwidth]{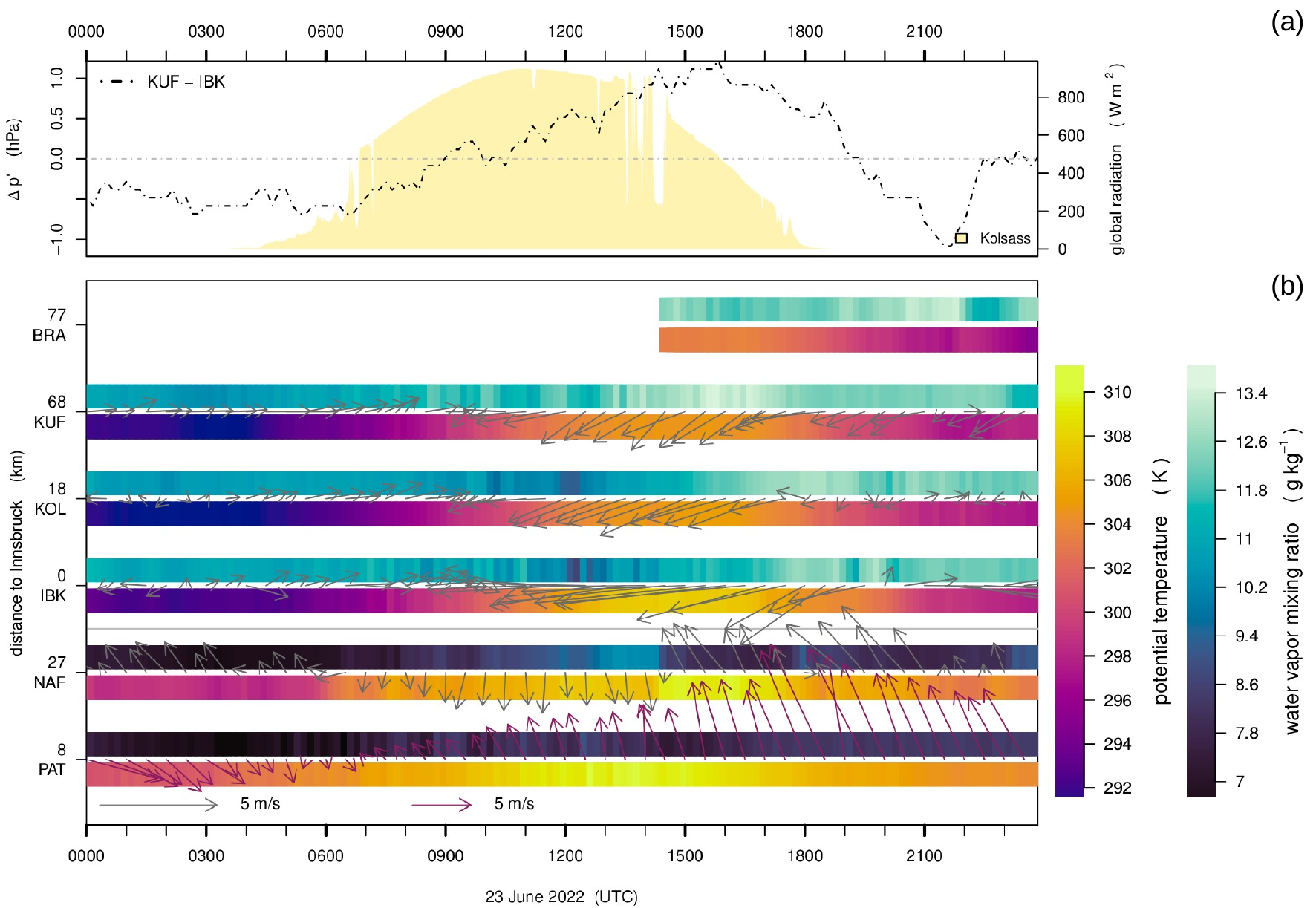}
    \caption{As in Fig.~\ref{fig4:awsaug} but from 0000~UTC 23~June to 0000~UTC 24~June 2022. Note that the mountain station Patscherkofel has a different reference arrow than the valley stations due to stronger winds at crest level.}
    \label{fig10:awsjun}
\end{figure}

The presence of foehn does not disrupt the valley wind pattern near the surface as indicated by the AWS observations (Fig.~\ref{fig10:awsjun}). Diurnal heating of the valley atmosphere leads on a local scale to a reversal of the pressure gradient and an associated reversal of the valley winds at all valley sites. Fluctuations in global radiation in the morning and afternoon indicate the presence of clouds.  At the mountain peak station Patscherkofel, winds change from a northerly to a southerly direction in the morning at about 0700~UTC, in agreement with a change in large-scale conditions (see~\ref{sec:synop2}). Afterwards, southerly winds continuously intensify and reach magnitudes of as much as 15~m~s$^{-1}$ in the late afternoon, a sign of the development of south foehn.

\begin{figure}
    \centering
    \includegraphics[width = \textwidth]{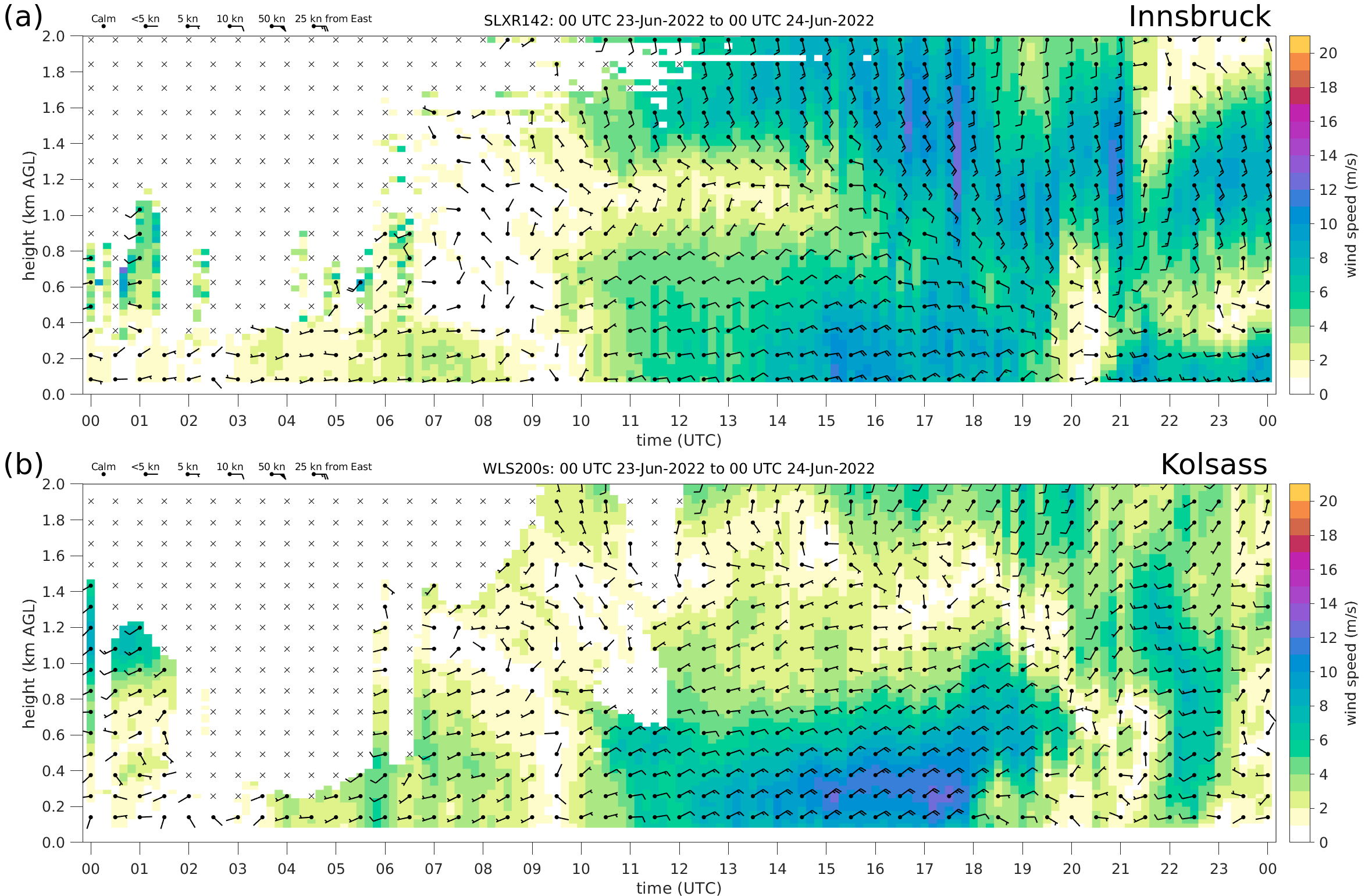}
    \caption{As in Fig.~\ref{fig5:adwl}a-b but on 23~June 2022.}
\label{fig11:2dwl}
\end{figure}

The interaction of the foehn and the valley winds, however, is apparent at the larger scales, as evidenced by Doppler wind lidar observations at IBK and KOL (Fig.~\ref{fig11:2dwl}). The upvalley flow starts at both locations between about 0900~and 1000~UTC, but further aloft moderate to strong southerly winds can be observed. These southerlies reach to lower heights at IBK than at KOL and thus keep the upvalley flow shallower at Innsbruck. They represent south foehn channeled through the Wipp Valley near Innsbruck and the Weer Valley near Kolsass. Foehn winds are stronger for the wider and deeper Wipp Valley, which is one of the most prominent Alpine foehn valleys \citep[e.g., ][]{Gohm2004}. In the late afternoon, the depth of the upvalley winds continuously decreases and the south foehn almost manages to penetrate to the surface at IBK. However, with the onset of near-surface cooling (see IBK in Fig.~\ref{fig10:awsjun}b), a CAP and associated downvalley winds establish below the foehn flow after 2030~UTC (Fig.~\ref{fig11:2dwl}a). These downvalley winds are stronger than pure nocturnal drainage winds (cf. Fig.~\ref{fig5:adwl}a and~\ref{fig11:2dwl}a) due to foehn-CAP interaction \citep[e.g., ][]{Haid2022,Umek2021}, and they are called pre-foehn westerlies. At KOL, south foehn is not able to descend as far as at IBK (Fig.~\ref{fig5:adwl}b). Instead, a deep layer of westerly (downvalley) flow establishes after 2100~UTC. These westerlies are most likely the result of an eastward deflected foehn current channelled by the Inn Valley and, again, not the result of pure nocturnal drainage winds. 

\subsubsection{Processes in the Weer Valley at Nafingalm}
\begin{figure}
    \centering
    \includegraphics[trim=25 40 130 80,clip,width = \textwidth]{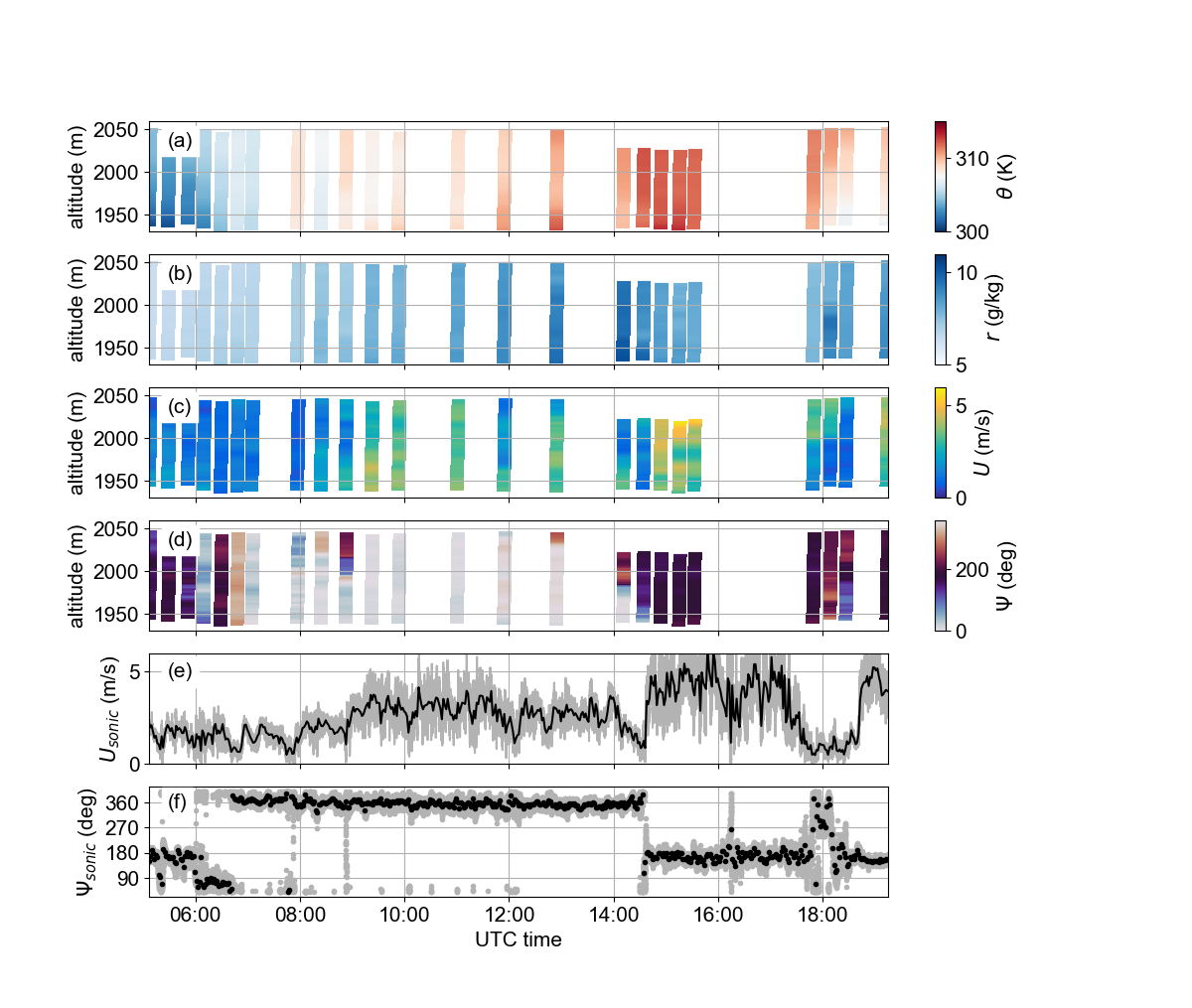}
    \caption{(a) Potential temperature, $\theta$, (b) water vapor mixing ratio, $r$, (c) wind speed, $U$, and (d) wind direction, $\Psi$, as measured by the fleet of UAS at NAF through vertical profiles up to 120~m AGL (2050~m above mean sea level) on 23~June 2022. Panels (e) and (f) show sonic anemometer measurements of wind speed and wind direction at 5~m AGL for the same period. The black markers show 15-minute averages while grey markers visualize the full resolution of 10~Hz. The main transition periods are in the morning at 0630~UTC, in the afternoon at 1430~UTC and in the evening at 1800~UTC as can be seen best from the wind direction changes.}
\label{fig12:swuf3d_overview}
\end{figure}

UAS profiling flights up to 120~m could give detailed insights into the role of tributary flows in the multi-scale interactions of foehn and valley wind systems beyond the capability of a surface network of AWS. On 23~June 2022, thermally-driven winds within NAF are indicated by quickly changing wind directions from southerly (downvalley) to northerly (upvalley) winds during the morning transition shortly after 0630~UTC and from northerly to southerly winds again during the evening transition at 1800~UTC (Fig.~\ref{fig12:swuf3d_overview}). During those transition periods, the surface sensible heat flux changes simultaneously with the wind direction which happens systematically also on other days (not shown). The almost instantaneous response of the winds at NAF to the change in the surface sensible heat flux may indicate that they are driven by local forcing, as in slope winds, rather than by the valley volume effect, as in valley winds. During the morning transition, potential temperature increases continuously, except for a short period of temporary cooling at 0830~UTC. 

\begin{figure}
    \centering
    \includegraphics[width = \textwidth]{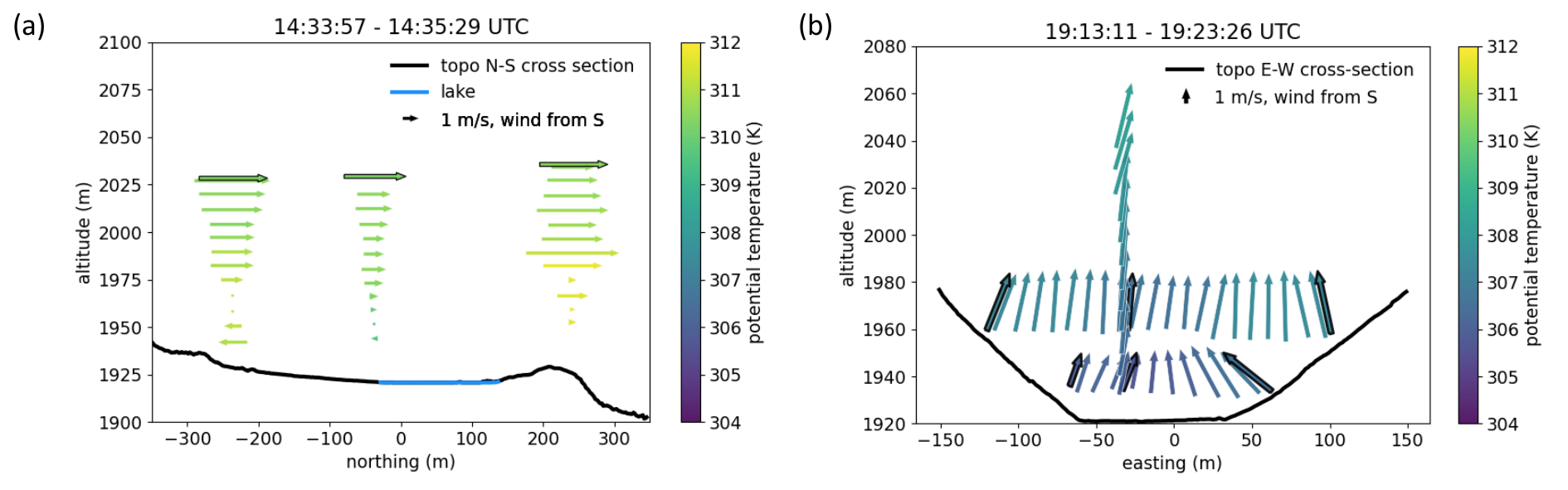}
    \caption{(a) Along-valley measurements from 1430~UTC and (b) cross-valley measurements from 1915~UTC performed by the UAS fleet at NAF on 23~June 2022. Arrow length shows the wind speed. Note the different meaning of arrow orientation in the panels: arrows in (a) represent the north-south component only, arrows in (b) indicate the total wind direction. Arrows with black edges are averaged values from a UAS hovering at one location for at least 30~seconds. Arrows without black edges in (a) are instantaneous values. Arrows of horizontal flights in (b) are averages of multiple cross-flight legs. Terrain height is indicated in (a) and (b) by a black line and the lake surface height in (a) by a blue line. Height scales are in meter above mean sea level. Note the difference in altitude range between panel (a) and (b).}
\label{fig13:swuf3d_vpro}
\end{figure}

In contrast to that, the afternoon boundary layer has features originating from the non-local south foehn starting at 0700~UTC at Patscherkofel (Fig.~\ref{fig10:awsjun}) but reaching NAF at approximately 1430~UTC. Between 1400 and 1500~UTC, the wind speed increases and turns downvalley to a southerly direction simultaneously with increasing temperatures and decreasing humidity as observed by ground observations and UAS measurements (Fig.~\ref{fig12:swuf3d_overview}). Three vertical profiles from synchronously operated UAS around 1430~UTC show similar but spatially heterogeneous patterns during this phase (Fig.~\ref{fig13:swuf3d_vpro}a). All profiles show strong southerly downvalley winds at the top of the profiles. However, these southerly winds do not entirely reach down to the ground at all three profiling sites. The northernmost profile only indicates southerly winds (right in Fig.~\ref{fig13:swuf3d_vpro}a) which also shows the warmest temperatures close to the ground, but at the southernmost profile (left in Fig.~\ref{fig13:swuf3d_vpro}a) local upvalley winds still persist below the south-foehn winds. The tendency of southerly winds establishing first at the northernmost profiling site could already be observed at 1410~UTC (not shown). Shortly after 1450~UTC, southerly winds then prevailed at all heights for all three profiling sites (not shown). A possible explanation of the low-level northerly winds in the southern part of the valley head at 1433~UTC might be progressive erosion of valley air masses by foehn air masses. However, alternative or superimposing effects such as flow separation and the formation of a lee side eddy or rotor \citep{Mursch1995} should be considered too. Further investigation is necessary to disentangle these multi-scale processes.

After 1730~UTC, the boundary layer temporarily cools, especially in the lowest 50-100~m AGL and winds at the ground almost completely cease, turning towards west at the measurement site north of the lake  (Fig.~\ref{fig12:swuf3d_overview}e, f). Around 1930~UTC, winds turn back to south and pick up again which is also observed by cross-flights with two UAS at 10~m and 40~m above the valley floor at a location just south of the lake (Fig.~\ref{fig13:swuf3d_vpro}b). Especially at 10~m, wind directions are from south-west above the western Weer Valley side wall, from south above the valley floor, and from south-east above the eastern Weer Valley side wall. The wind component pointing away from the slopes might be an indication of downslope winds. At least for the easternmost point of the cross-flight of the lower UAS, the minimum distance of the UAS to the slope was only 5\,m. At that location, a gully lead to the valley floor, potentially channeling and thickening nocturnal drainage flow. Under this assumption, the UAS provide important measurements to quantify how the mesoscale southerly downvalley wind gets mixed with local nocturnal drainage winds. More spatial measurements along the slope and at different heights will be necessary in the future to get a full picture of the scale interactions.

\subsection{Further tested measurement setups}\label{sec:further_measurements_tested}
Some instruments were not operated continuously or had data gaps and therefore could not be included in the analysis of the previous two events. Nevertheless, their importance for the TOC is described in this section for 17~July 2022 which was a synoptically undisturbed day with mainly clear sky (not shown). 

\begin{figure}
    \centering
    \includegraphics[width = \textwidth]{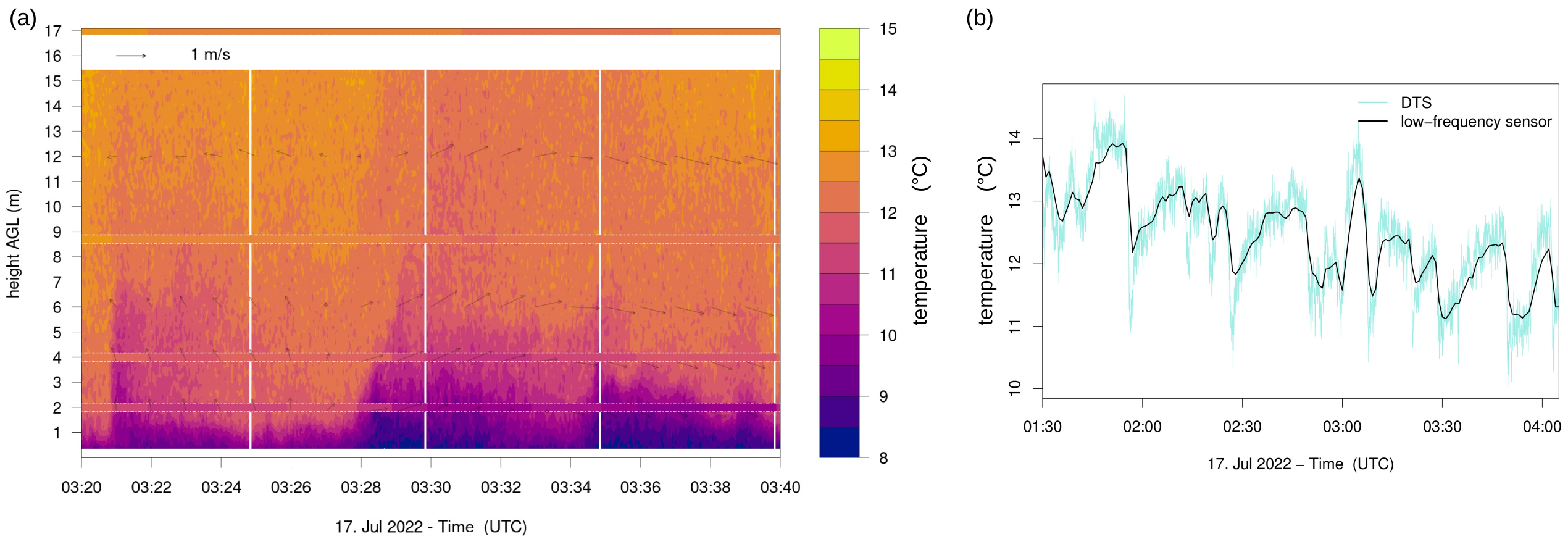}
    \caption{DTS measurements conducted at KOL on 17~July 2022: (a) Time-height-diagram of DTS temperatures together with low-frequency (1~min) temperatures ventilated sensors at 2, 4, 8.7, and 17~m~AGL (between the white horizontal lines), and arrows for wind speed and direction at 2,~4,~6, and 12~m~AGL. (b) Time series of DTS measurements at 4~m AGL together with temperature measured by a low-frequency sensor for a selected period of 2.5 hours. }
    \label{fig14:DTS}
\end{figure}

The main advantage of DTS is its fine-scale resolution revealing cold-air layers below the lowest point observation (Fig.~\ref{fig14:DTS}a, dark purple below 2~m AGL) and the almost wave-like changing depth of the cold-air layer after 03:35~UTC around 3~m AGL (Fig.~\ref{fig14:DTS}a) which can not be captured by point observations. Further, due to the higher temporal resolution, DTS can detect features on smaller temporal scale than the point observations (Fig.~\ref{fig14:DTS}b). Accordingly, due to the high spatial and temporal resolution the DTS technique offers a very detailed observation of the near-surface layer beyond the capability of the ventilated low-frequency sensors. The DTS observations can also be used to define the morning and evening transitions. For future studies, a much taller vertical DTS deployment could determine the height of the boundary layer  \citep{Fritz2021}. Other studies show that point observations miss, for example, small-scale air layers of different stability or transient features which can only be resolved by distributed observations like DTS \citep[e.g. ][]{Peltola2021, Pfister2021, Pfister2021a, Schilperoort_2022}, highlighting the potential necessity of DTS data for answering certain questions regarding boundary layer dynamics.

Nevertheless, the DTS experiment also had some issues mainly due to the mounting of the fiber-optic cable. Anything touching the fiber will create an artifact so that data at the bottom and top of the array had to be discarded, shortening the vertical extent of DTS measurements. Radiation artifacts from shading by the tower are evident in the morning from 0400~until 0930~UTC (not shown). Artifacts from solar radiation are a well-known limitation for DTS \citep{Sigmund2017}.

\begin{figure}
    \centering
    \includegraphics[width = \textwidth]{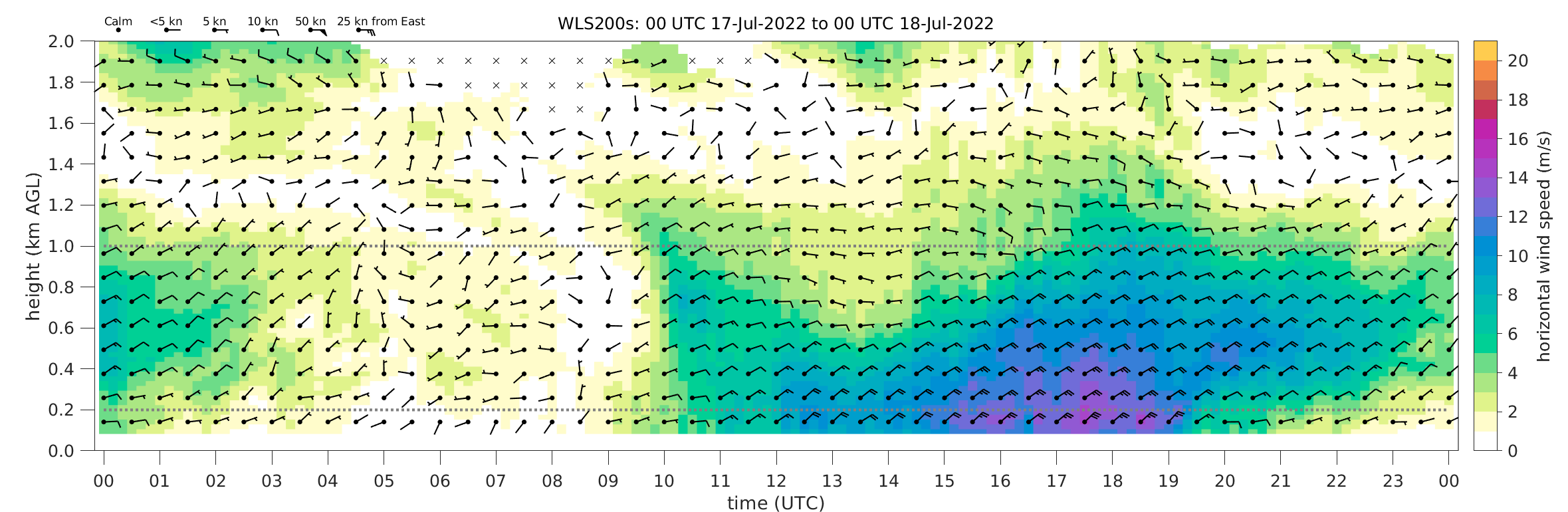}
    \includegraphics[width = \textwidth]{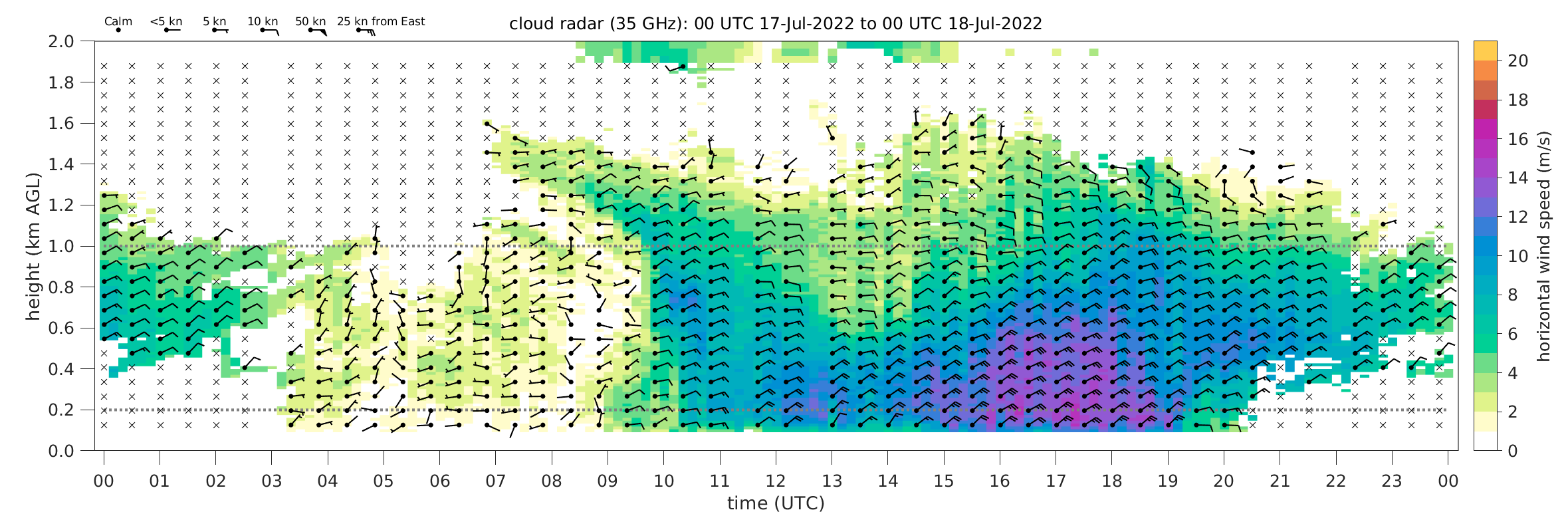}
    \unitlength1mm
    \begin{picture}(0,0)
      \put(-90,115){\large{(a)}}
      \put(-90,60){\large{(b)}}
    \end{picture}
    \caption{Time-height diagram of 10-min average horizontal wind speed (color shading) and horizontal wind direction (wind barbs) measured on 17 July 2022 by (a) the Doppler wind lidar WLS200s-115 and (b) the RPG FMCW cloud radar at 35 GHz, both located at KOL. Crosses mark missing data due to a too low signal-to-noise ratio.}
    \label{fig15:radar_vs_lidar}
\end{figure}

\begin{figure}
    \centering
    \includegraphics[width = .6\textwidth,clip,viewport=100 260 480 560]{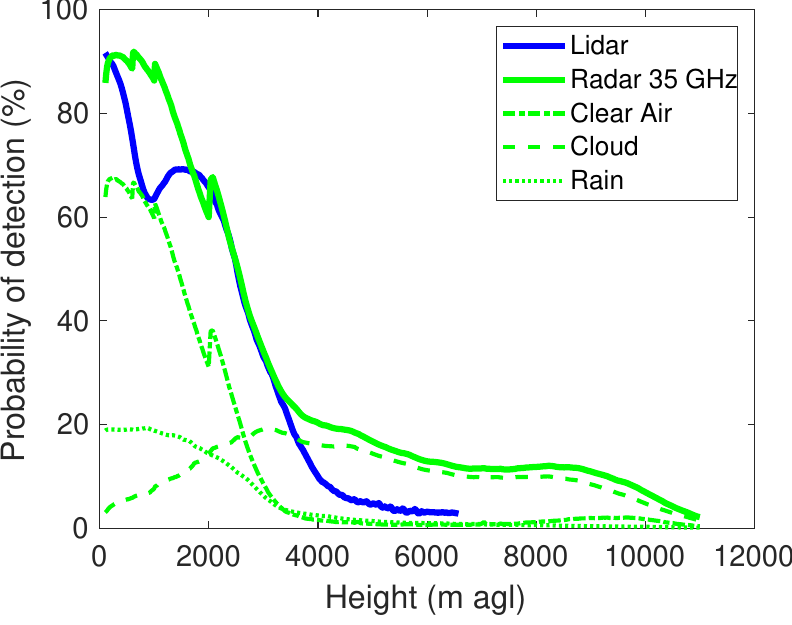}
    \caption{Probability of detection of vertical wind speed by the WLS200s-115 lidar (operating 85~days) and the RPG FMCW cloud radar at 35 GHz (operating 54~days) for vertical pointing measurements. A simple scatterer classification system (Section~\ref{sec:further_measurements_tested}) determines whether the cloud radar measurements were taken in clear air, clouds, or rain. }
    \label{fig16:propability_detection}
\end{figure}

Another objective of this test campaign was to determine how well a cloud radar is suited to perform wind measurements compared to a Doppler wind lidar. The focus here is on the analysis of data availability and vertical range. The RPG FMCW dual frequency 94/35~GHz scanning cloud radar was installed close to the WLS200s-115 Doppler wind lidar in KOL (Section~\ref{sec:KOL}). The analysis is based on cloud radar data covering 54~days out of a total of 100 days (considerable data gaps were caused by technical problems which are now identified) and on WLS200s-115 data from 85~days out of 101~days. Only results from the cloud radar measurements at 35~GHz are shown (Fig.~\ref{fig15:radar_vs_lidar} and~\ref{fig16:propability_detection}), as those are more reliable than the measurements at 94~GHz.

During a day without precipitation the lidar measurements of horizontal wind reach higher altitudes and indicate slightly lower wind speeds than the cloud radar measurements (Fig.~\ref{fig15:radar_vs_lidar}). There is a certain degree of freedom to accept or reject measurements on its signal-to-noise-ratio and the consistency around the azimuths. Such criteria as well as spatial and temporal resolutions have not been aligned between the two systems. 
Nevertheless, to compare the measurement ranges of both instruments, the probability of detection is calculated from the vertical pointing measurements. It is well known that cloud radars, in opposite to lidars, are able to measure in clouds and during (light) precipitation. Our interest was how far cloud radar clear air echoes can provide wind information. For this purpose, the echoes of the cloud radar are identified by a simple scatterer classification system: 
\begin{itemize}
  \item Scatterer falling faster than 2 m/s ($w<-2$\,m/s) are identified as rain.
  \item Scatterer with a Slanted Linear Depolarization Ratio (SLDR) higher than -25\,dB which are not or hardly sedimenting ($w>-0.5$\,m/s) are identified as clear air echoes as well as scatterer with an absolute differential reflectivity of more than 0.5\,dB ($|\mbox{ZDR}|>0.5$\,dB).
  \item All other non-empty echoes are identified as cloud particles.
\end{itemize}
A visual inspection proves that this classification scheme is sufficient for a statistical analysis.  The scatter plot of the lidar scatterer speed against the backscatter signal (not shown) indicates that lidar wind measurements can be accepted down to a backscatter of $-70$\,dB. 

With these criteria the probability of detection as a function of the measurement height shows comparable results for lidar and cloud radar but only up to about 3500\,m (Fig.~\ref{fig16:propability_detection}). In the lowest 2\,km the cloud radar has a slightly higher availability of wind data. The number of successful wind measurements by the cloud radar due to clear air echoes does not quite reach that of lidar measurements, but additional measurements can be made in clouds and especially in rain. The step in cloud radar availability at 2000\,m is caused by the change of the ``chirp'' and the reduction of radial resolution. Above roughly 3500 m, measurements within clouds are almost exclusively available from the cloud radar with a probability of 20~\% while the range of the lidar is limited here.

\section{Resolvable scales and gained experiences for the TOC \label{sec:scales_and_lessons}}
\begin{figure}
    \centering
    \includegraphics[width = \textwidth]{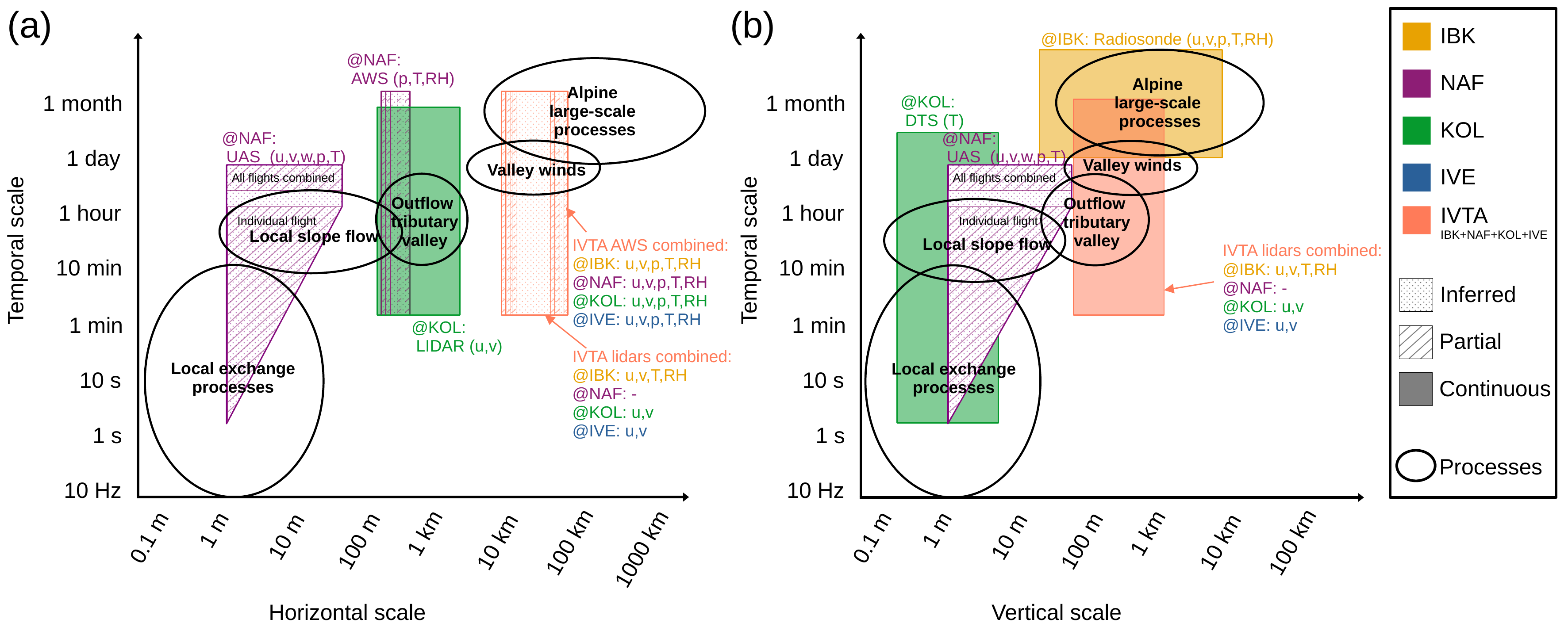}
    \caption{Schematic space-time diagram of resolvable scales during TEAMx-PC22 divided into (a) horizontally and (b) vertically resolvable scales. The Inn Valley target area (IVTA) and corresponding sub-target areas are indicated by colours. The filling of areas categorizes if the resolvable scale is inferred, partial, or continuous (cf. Section~\ref{sec:scales_and_lessons}). A selection of processes and motions under investigation during TEAMx is also added (bold circles and text) to illustrate which of them were actually resolvable and at which location. All resolvable scales have a short description of which instruments were used (AWS: automatic weather station; lidar: remote sensing with light detection and ranging; UAS: uncrewed aerial system; DTS: distributed temperature sensing; radiosonde) and which parameters were retrieved (T: temperature; RH: relative humidity; p: pressure; u,v: horizontal wind speed and direction; w: vertical wind speed). }
    \label{fig17:resolved_scales}
\end{figure}

During TEAMx-PC22 processes on a range of scales were observed by different measurement techniques. Some of the measurement techniques resolve processes from one location in time, others have a certain temporal and spatial resolution. Further, a network of point-observations can be utilized for observing a process in space, but the spacing is usually less regular and coarser than lidar measurements (depending on the network setup of course). An attempt is made to visualize the resolvable scales from using and combining different measurement techniques in a schematic space-time diagram (Fig.~\ref{fig17:resolved_scales}). Spatial scales were divided into horizontal (Fig.~\ref{fig17:resolved_scales}a) and vertical (Fig.~\ref{fig17:resolved_scales}b) scales. The diagram highlights which processes were resolvable at each sub-target area and can be further investigated in future publications or utilized for planning the TOC. 

The resolvable scales in time and space were determined using the analogy of a (fast) Fourier transform: the lower boundary is twice the highest spatial/temporal sampling rate and the upper boundary is half of the fully observed spatial/temporal domain. For example, a vertically staring lidar with a sampling rate of 25~m observing from a few tens of meter up to 2~km AGL and a temporal sampling rate of 1~min measuring for the full span of 2~months is shown as a vertical resolvable spatial scale from 50~m to 1~km and a resolvable temporal scale from 2~min up to 1~month  (Fig.~\ref{fig17:resolved_scales}b, opaque orange area). Physical processes occurring with a scale smaller or larger than these limits cannot be unambiguously resolved; processes at a finer scale will be aliased while those at a larger scale cannot be separated from secular trends. 

Further, three categories of resolvable scales were determined based on how typical scale-analysis techniques could be applied in space and time, like a Fourier transform or continuous wavelet transform. Continuously resolvable scales are those with regular sampling rate in time and space, meaning a Fourier transform could be applied along both the space and time dimensions independently (Fig.~\ref{fig17:resolved_scales}, opaque area). Examples of continuously resolvable scales would be some lidar configurations and DTS. Partially resolvable scales are those without regular sampling in time and space simultaneously, meaning a Fourier transform could not be applied for a single location along the time dimension or a single time along the space dimension. However, these observations are rich and can still yield process scales, albeit without the benefit of typical analytical techniques such as a Fourier transform. Instead, analysis relies on human intuition, statistical models, and/or physical models to provide the missing details. Examples of these observations include non-stationary UAS, which depending on the flight pattern provide repeated observations of the atmosphere at the same place with irregular spacing in time (Fig.~\ref{fig17:resolved_scales}a, hatched purple areas). The plot only visualizes a UAS continuously moving back and forth along a vertical or horizontal transect which was repeated several times during a day and for several days. Inferred scales are those where a Fourier transform could only be applied along a single dimension and some sort of statistical or physical model would be needed to infer the scales along the other dimension. Examples of inferred scales are the network of AWS or lidars within the IVTA, where the observations are regular in time but the spacing between stations is irregular with large gaps (Fig.~\ref{fig17:resolved_scales}, orange dotted area). The actual spacing between stations is shown as partially resolvable scale (hatched area) within the inferred resolvable scale (dotted area). A single point observation (with regular observation in time) or its footprint was excluded from the diagram since  the observation does not have a spatial sampling rate which is continuous in time nor a temporal sampling rate which is continuous in space as it depends on wind speed as well as direction from where the air is transported from (when assuming Taylor's frozen turbulence to be valid). Atmospheric processes which were desired to be resolved were added as bold ellipses.   

The setup of each sub-target area made it possible to resolve at least one atmospheric process in great detail. Within the sub-target area IBK, the combination of lidars and radiosondes could resolve the local boundary layer vertically in detail revealing the evolution of the nighttime CAP and daytime CBL. The sub-target area NAF was well equipped to study the local thermally-driven winds and their interaction with south foehn with UAS and a network of stations. Within the sub-target area KOL, the lidars were able to investigate the outflow from the tributary Weer Valley, and within the sub-target area IVE, the Inn Valley exit-jet was observed in a more detailed way than any other study so far. Furthermore, this conceptual model highlights the importance of a network of sub-target areas during TEAMx-PC22. Due to AWS and lidar measurements  within each sub-target area, valley winds in the Inn Valley could be investigated in great detail.

Additionally, the space-time diagram identified observational gaps. Due to sparse spacing of AWSs and lidars some gaps within the IVTA can be seen when looking at the horizontally inferred resolution and space between the vertical bars (Fig.~\ref{fig17:resolved_scales}~a, dotted areas). The space could be minimized and thus the resolution improved by having another location between KOL and IVE potentially providing more details about the onset of valley winds. For investigating the interaction between south foehn, valley winds, and slope winds within the Weer Valley a higher spatial resolution in combination with covering a bigger spatial domain is needed to resolve processes from the smallest (slope winds) to biggest scales (south foehn). Smaller drones with higher temporal resolution could be deployed with a special flight pattern along the Weer Valley side walls close to the surface to investigate the rather weak slope winds even within small gullies. To connect the observations within NAF to larger scales, additional measurements aloft with UAS or other instruments, over the crest, and potentially along the Weer Valley down to KOL, would be beneficial. Further, the determination and quantification of the valley exit-jet requires additional radiosondes within IVE to prove the development of a hydraulic jump and the associated lowering of isentropes as described by \cite{Zangl2004}.  

The presented space-time diagram of resolved scales and targeted processes can be utilized not only after a field campaign. Planning and coordination of field campaigns would benefit from building these time-space maps beforehand and explicitly describing the resolvable scales and identifying the associated gaps. The gained knowledge during TEAMx-PC22 will be utilized for planning the TOC.

Other lessons learned from TEAMx-PC22 are the need for high-precision pressure sensors within the sub-target area IVE. Such sensors could provide the possibility to distinguish between the forcing from the hydrostatic pressure differences resulting from the vertically integrated temperature difference between the valley and the foreland, and the dynamic pressure variations resulting from flow acceleration and deceleration along the valley exit due to the terrain constriction, the hydraulic jump, and the transition from the Inn Valley to the Bavarian foreland.

All chosen sites were appropriate locations and can be utilized for the TOC. This was especially critical for the sub-target area NAF. Due to its remoteness, measurement techniques that have high power requirements, such as lidars, were not possible during the TEAMx-PC22 and will pose a great logistic challenge for the TOC. Nevertheless, the deployed UAS in combination with the network of near-surface stations enabled us to study scale interactions. 

Due to delays or technical difficulties, not all instruments were running simultaneously during TEAMx-PC22. Most remote sensing instruments were mainly available in August and September while DTS and UAS were mainly operated in June and July. With longer intensive observation periods and more formalised communication and coordination, TEAMx-PC22 would have led to more overlap of data. The use of the resolvable scales diagram may help communicate the role each system has during coordinated intensive observation periods and assist in targeting the deployment of limited resources.

For the TOC it is desirable to close all observational gaps, to improve the tested setup, to extend observations temporally and spatially, and to have coordinated cross-Alpine flights by instrumented research aircraft of different types equipped with in-situ and remote sensing instrumentation to investigate processes up to the Alpine scale.

\section{Summary of the TEAMx pre-campaign \label{sec:summary}}
During TEAMx-PC22 all tested field sites and setups successfully contributed to create a multi-scale dataset which will lead to new insights into the MoBL as well as meso- and microscale flows over Alpine terrain. The presented case studies show a large variety of phenomena that are relevant for transport and exchange processes over mountains. They also demonstrate the interaction between local, regional, and synoptic winds and thus the scale interaction within complex terrain. Furthermore, a new measurement site suitable for the TOC could be determined (sub-target area NAF). 

During the first case study with synoptically undisturbed conditions, the evolution of the MoBL was observed and described in detail within the sub-target area IBK. The transition to a CBL and the onset of upvalley winds was observed almost simultaneously throughout the IVTA. However, the time and strength of the maximum flow as well as the onset of the downvalley flow were station-dependent and should be analyzed in more detail. The interaction of the outflow from a tributary (Weer Valley) with the flow in the main valley (Inn Valley) was captured in the sub-target area KOL through a careful configuration of observational systems. The outflow usually experienced a deflection towards the direction of the flow in the Inn Valley in the late evening and early morning, while during the night the outflow was meandering due to the weaker flow in the main valley.  A more systematic analysis of the outflow from tributary valleys should lead to new insights into scale interactions. Within the sub-target area IVE, a valley exit-jet was observed during the night leading to spatially heterogeneous warming near the surface. 

The second case study illustrated multi-scale interactions between south foehn and valley winds in the Inn Valley and its tributary Weer Valley. While near the surface valley winds established, an interaction with the south foehn at higher levels was location dependent and generated a heterogeneous wind field throughout the IVTA. During the day, the south foehn penetrated further down at IBK than at KOL. In KOL, the foehn manifested itself as an eastward deflected current. Furthermore, the south foehn also reached the sub-target area NAF during the early afternoon with a gradual transition from upvalley winds to foehn-induced downvalley winds. Also, during the night the wind field was heterogeneous due to flow down the Nafingalm side walls while the downvalley flow was most likely accelerated by the weaker but still prevailing foehn winds.

TEAMx-PC22 can be considered a success as new instruments and measurement techniques were successfully tested, new measurement sites for the TOC were determined, and last but not least a unique dataset was collected which is publicly available on the TEAMx-PC22 Zenodo community\footnote{\label{foot:zenodo}TEAMx-PC22 Zenodo community: \url{https://zenodo.org/communities/teamx-pc22/}}. 
The goal for the TOC is to close the observational gaps  and extend the measurements in the IVTA to different processes, such as turbulent exchange, nocturnal drainage winds, and air quality, and to embed the observations in the IVTA into the large-scale, cross-Alpine design of the TOC as outlined in \cite{Serafin2018}. To this end, the TEAMx-PC22 dataset can be used for further detailed process studies, for testing and demonstrating new observational methods, and for NWP and LES model verification as well as NWP data assimilation.

\section*{Data availability}
The TEAMx-PC22 campaign took place from mid May until end of September 2022, however, data availability varies between instruments due to delays or technical difficulties with the instruments. A full summary with all dates is given in Table~\ref{tab:instruments}.  Datasets gathered during TEAMx-PC22 are publicly available online on  the TEAMx-PC22 Zenodo community (\url{https://zenodo.org/communities/teamx-pc22/}) or on request from the instrument operators. The TEAMx-PC22 Zenodo community also includes an introductory video clip \citep{Gohm2023_videoClip} for KOL and NAF. 
For the sub-target area IBK, radiosonde measurements are available online\footnote{\label{foot:radiosonde}\url{https://weather.uwyo.edu/upperair/bufrraob.shtml}}, while the Doppler wind lidar data sets (SL88, SLXR142)  \cite{Gohm_2023_lidars_IBK}. 
For the sub-target area KOL, horizontal 10-min winds and 1-s vertical winds from the cloud radar \citep{Handwerker2023_Cloud_radar_horizontal_winds, Handwerker2023_Cloud_radar_vertical_winds}, 1-s DTS data, 30-min flux data at two levels, and 1-min data including wind profile, temperature profile, radiation, and pressure measurements \citep{Pfister2023_InnDEX22}, Doppler wind lidar VAD products \citep{Babic2023_VAD_windcube} and the horizontal wind field retrieved from measurements with two Doppler wind lidars \citep{Babic2023_Radial_velocity} as well as animations of the tributary outflow \citep{Babic2022} are available. 
Sub-target area NAF features datasets from a network of temperature and humidity loggers \citep{Gohm_2023_Hobo_Nafingalm} and an AWS \citep{Gohm_2023_AWS_Nafingalm}, as well as a UAS dataset including vertical profiles and fixed-point measurements \citep{wildmann:23}. 
For the sub-target area IVE, data of the AWS network \citep{Paunovic2023a}, radiosonde launches and drone measurements at Brannenburg \citep{Paunovic2023b}, ceilometer measurements at Kufstein \citep{baumann_stanzer_2023_ceilo}, and Doppler wind lidar measurements at Kufstein  \citep{baumann_stanzer_2023_lidar} and Brannenburg \citep{Leinweber2023} are provided.

\section*{Acknowledgments}
This research has been supported by the HORIZON EUROPE European Research Council (grant no. 101040823) and by the Austrian Science Fund (FWF grant ESP~214-N and V~791-N). Special thanks to Philipp Vettori and Friedrich Obleitner for their excellent support during TEAMx-PC22. We thank Austro Control GmbH, especially Andreas Lanzinger, for launching additional radiosondes at Innsbruck Airport. Further, we appreciate KIT/IMK-IFU for lending us the DTS measurement device, the DWD Mobile Measurement Unit for installing and operating AWS, radiosondes and an instrumented drone, as well as KITcube technicians for their support during all phases of TEAMx-PC22. We also want to thank all landowners at  Nafingalm, Fritzens, Terfens, Flintsbach, Oberaudorf, Kiefersfelden, Hohe Asten, as well as the Kläranlage Brannenburg and the power provider TINETZ/TIWAG, Stadtwerke Schwaz. Lastly, we want to thank Meteo Swiss for providing COSMO-1 analysis and forecast data.

\bibliographystyle{apalike}
\setcitestyle{authoryear,open={(},close={)}}
\bibliography{bibliography_TEAMx_PC22}   
 \clearpage
\section*{Attachments}
\fontsize{9}{6}\selectfont

\captionsetup{width=.95\textheight}
\begin{landscape}
\begin{longtable}[l]{p{2cm}p{1.7cm}p{1.2cm}p{4cm}p{2.2cm}p{2cm}p{2.5cm}p{2cm}p{2.8cm}}
\caption{\fontsize{9}{6}{Table of instruments within the Inn Valley target area during TEAMx-PC22. The instruments are sorted by sub-target area. The measured variables are air temperature (T), relative humidity (H), pressure (P), two horizontal wind components (U, V), vertical wind component (W), radial velocity (Vr), radiation (R), precipitation (RR), soil parameters (S), backscatter (B), reflectivity (Z), elevation angle (EL), azimut angle (AZ), auxiliary parameters (AUX).}\label{tab:instruments}}\\ \hline
\textbf{Site name} & \textbf{Coordinates \linebreak (lat/lon)} & \textbf{Site \linebreak Altitude  (m ASL)} & \textbf{Instruments} & \textbf{Instrument \linebreak abbreviation} & \textbf{Measured variables} & \textbf{Description measurement method} & \textbf{Institution} & \textbf{Deployment period} \\ \hline
\endfirsthead
\multicolumn{9}{c}%
{{\bfseries Table \thetable\ continued from previous page}\vspace{2mm}} \\ \hline
\textbf{Site name} & \textbf{Coordinates \linebreak (lat/lon)} & \textbf{Site \linebreak Altitude  (m ASL)} & \textbf{Instruments} & \textbf{Instrument \linebreak abbreviation} & \textbf{Measured variables} & \textbf{Description measurement method} & \textbf{Institution} & \textbf{Deployment period}  \\ \hline
\endhead
\multicolumn{9}{l}{\cellcolor[HTML]{D9D9D9}\textbf{Sub-target area IBK}} \\
\textbf{Innsbruck University} & 47.264083 N 11.384986 E & 575 & Purple Pulse Raman lidar & PPL & T, H, B & vertical profile (stare) & KIT/IMK-IFU & 09~Aug-25~Sep 2022 \\
\textbf{Innsbruck University} & 47.264083 N 11.384986 E & 575 & Halo Photonics Streamline Doppler wind lidar (SN 88) & SL88 & Vr, W, B & vertical profile (stare) & ACINN & 11~Aug-02~Oct 2022 \\
\textbf{Innsbruck University} & 47.264310 N 11.385290 E & 613 & Halo Photonics Streamline XR Doppler wind lidar (SN 142) & SLXR142 & Vr, U, V, W, B & PPI \at EL=70\textdegree~ & ACINN & semi-permanent \\
\textbf{Innsbruck University} & 47.264310 N 11.385290 E & 613 & TAWES weather station and IAO turbulence tower & AWS+turb & U, V, T, H, P, R, RR, S, AUX & near-surface & ACINN \& GeoSphere Austria & permanent \\
\textbf{Innsbruck Airport} & 47.259800 N 11.355340 E & 579 & Radiosonde Vaisala RS41 & RS & U, V, T, H, P & vertical profile & Austro Control & 14 days between 23~Aug-24~Sep 2022, 6-hourly \\
\multicolumn{9}{l}{\cellcolor[HTML]{D9D9D9}\textbf{Sub-target area KOL}} \\
\textbf{Kolsass} & 47.305290 N 11.622231 E & 546 & Leosphere Windcube WLS200s (SN 115) & WLS200s-115 & Vr, U, V, W, B & PPI and DBS \at EL=70\textdegree~ & KIT/IMK-TRO & 08~Jun-17~Sep 2022, gap: 16~Jun,02~Jul-16~Jul 2022 \\
\textbf{Kolsass} & 47.305290 N 11.622231 E & 546 & Leosphere Windcube v2.1 (SN 1489) & WCv2.1 & U, V, W & DBS \at EL=62\textdegree~ & KIT/IMK-TRO & May-Sep 2022 \\
\textbf{Kolsass} & 47.305290 N 11.622231 E & 546 & RPG FMCW cloud radar & FMCW-W/Ka & Vr, U, V, W, Z, AUX & 10 min pattern: PPI \at EL=70\textdegree~; vertical profile (stare) & KIT/IMK-TRO & 18~May-26~Aug 2022, several data gaps \\
\textbf{Kolsass} & 47.305290 N 11.622231 E & 546 & i-Box turbulence tower & AWS+turb & U, V, W, T, H, P, R, RR, AUX & near-surface, multi-level & ACINN & permanent \\
\textbf{Kolsass} & 47.305290 N 11.622231 E & 546 & Fiber-optic distributed temperature sensing & DTS & T & vertical profile & ACINN & 08~Jun-14~Jun, 28~Jun-17~Jul 2022 \\
\textbf{Fritzens} & 47.321660 N 11.639410 E & 612 & Leosphere Windcube WLS200s (SN 159) & WLS200s-159 & Vr, (U, V), B & PPI \at EL=0\textdegree~ & KIT/IMK-TRO & 18~May-17~Sep 2022, data gap: 02 -16~Jul 2022 \\
\textbf{Terfens} & 47.309358 N 11.613750 E & 610 & Leosphere Windcube WLS200s (SN 124) & WLS200s-124 & Vr, (U, V), B & PPI \at EL=0\textdegree~ & KIT/IMK-TRO & 18~May-17~Sep 2022 \\
\multicolumn{9}{l}{\cellcolor[HTML]{D9D9D9}\textbf{Sub-target area NAF}} \\
\textbf{Nafingalm\linebreak (valley floor)} & 47.212400 N  11.713200 E & 1921 & SWUF-3D UAS & UAS & U, V, W, T, H, P & vertical profiles, horizontal legs, multi-point hover, up to 120~m AGL & DLR & 20-28~Jun 2022 \\
\textbf{Nafingalm\linebreak (north of lake)} & 47.215140 N 11.712630 E & 1928 & Sonic Anemometer Metek USA Class A & USA & U, V, W, T & near surface & DLR & 20-28~Jun 2022 \\
\textbf{Nafingalm\linebreak (north of lake)} & 47.215140 N 11.712630 E & 1928 & Automatic Weather Station & AWS & U, V, T, H, P, R, RR, S & near-surface, multi-level & ACINN & 15~Jun-12~Sep 2022 \\
\textbf{Nafingalm\linebreak (north of lake)} & 47.215140 N 11.712630 E & 1928 & HOBO MX2302 \&  MX2201 & T/H logger & T, H & near-surface, multi-level & ACINN & 16~Jun-12~Sep 2022 \\
\textbf{Nafingalm\linebreak (south of lake)} & 47.212760 N 11.713030 E & 1921 & HOBO MX2302 \& MX2201 & T/H logger & T, H & near-surface, multi-level & ACINN & 16~Jun-12~Sep 2022 \\
\textbf{Nafingalm\linebreak (inside  lake)} & 47.213600 N 11.712430 E & 1921 & HOBO MX2201 & T logger & T & in-lake, multi-level & ACINN & 16~Jun-12~Sep 2022 \\
\textbf{Nafingalm\linebreak (slope)} & 47.208150 N 11.721200 E & 2241 & HOBO MX2302 \& MX2201 & T/H logger & T, H & near-surface, multi-level & ACINN & 16~Jun-12~Sep 2022 \\
\textbf{Nafingalm \linebreak (crest)} & 47.202940 N 11.730160 E & 2531 & HOBO MX2302 & T/H logger & T, H & near-surface & ACINN & 16~Jun-12~Sep 2022 \\
\multicolumn{9}{l}{\cellcolor[HTML]{D9D9D9}\textbf{Sub-target area IVE}} \\
\textbf{Brannenburg} & 47.741547 N 12.122187 E & 456 & Halo Photonics Streamline XR (SN 172) & SLXR172 & Vr, U, V, W, B, AUX & PPI \at EL=35\textdegree~ \linebreak RHI \at EL=3\textdegree-51\textdegree, \linebreak AZ=150\textdegree~-160\textdegree~ & DWD & PPI (CSM): 15~Jun-26~Jul, 13~Aug-19~Oct 2022; RHI and PPI (step-stare): 27~Jul-12~Aug 2022 \\
\textbf{Brannenburg} & 47.742360 N 12.121712 E & 456 & LTS2000, HMP45d, Gill, mSonic3, PTB330, CNR4, Pluvio, HFP01SC & AWS & U, V, W, T, H, P, R, RR, S, AUX & near-surface & DWD & 24~Jun 2022-end of TOC \\
\textbf{Brannenburg} & 47.742376 N 12.121785 E & 456 & Radiosonde Vaisala RS41 & RS & U, V, T, H, P & vertical profile & DWD & 18-19~Jul 2022 \\
\textbf{Brannenburg} & 47.742840 N 12.121808 E & 456 & Drone DJI Mavic pro, iMET-XQ2 & UAS & T, H, P & vertical profile, up to 120~m AGL & DWD & 18-19~Jul 2022 \\
\textbf{Flintsbach} & 47.728401 N 12.127413 E & 468 & LTS2000, HMP45d, Thies & AWS & U, V, T, H & near-surface & DWD & 02~Jun 2022-end of TOC \\
\textbf{Hohe Asten} & 47.703476 N 12.115288 E & 1226 & LTS2000, HMP45d, Thies & AWS & U, V, T, H & near-surface & DWD & 02~Jun 2022-end of TOC \\
\textbf{Kiefersfelden} & 47.607147 N 12.202737 E & 473 & LTS2000, HMP45d, Thies, PTB110 & AWS & U, V, T, H, P & near-surface & DWD & 05~Jul 2022-end of TOC \\
\textbf{Oberaudorf} & 47.654282 N 12.180008 E & 466 & LTS2000, HMP45d, Thies & AWS & U, V, T, H & near-surface & DWD & 02~Jun 2022-end of TOC \\
\textbf{Kufstein} & 47.575280 N 12.162780 E & 490 & Ceilometer, Vaisala CL51 & CL51 & B, AUX & vertical profile (stare) & GeoSphere Austria & 01~Jun 2022-permanent \\
\textbf{Kufstein} & 47.575280 N 12.162780 E & 490 & Metek Wind Ranger 200 (SN 13) & WR13 & U, V, W, B & PPI \at EL=80\textdegree~ & GeoSphere Austria & 17~Aug 2022-03 Oct 2022 \\
\textbf{Kufstein} & 47.575280 N 12.162780 E & 490 & TAWES weather station & TAWES & U, V, T, H, P, R, RR, S & near-surface & GeoSphere Austria & permanent
\end{longtable}
\end{landscape}

\end{document}